\documentclass[aps, prd, letterpaper, 12pt, nofootinbib, superscriptaddress, longbibliography, notitlepage, floatfix]{revtex4-2}
\usepackage[utf8]{inputenc}
\usepackage{amsmath,amssymb,amsfonts}
\usepackage{mathrsfs}
\usepackage{xcolor}
\usepackage{graphicx} 
\usepackage{subfigure}
\usepackage{slashed}
\definecolor{hgreen}{rgb}{0,.7,0}
\definecolor{hred}{rgb}{.7,0,0}
\definecolor{hblue}{rgb}{0,0,.7}

\usepackage[colorlinks=true,
linkcolor=hblue,
citecolor=hgreen,
filecolor=hblue,
urlcolor=hred]{hyperref}

\allowdisplaybreaks

\begin{document}

%%%%%%%%%%%%%%%%%%%%%%%%%%%%%%%
\title{Updated Constraints from Electric Dipole Moments in the MSSM with R-Parity Violation}
%%%%%%%%%%%%%%%%%%%%%%%%%%%%%%%

\author{Wolfgang~Altmannshofer}
\email{waltmann@ucsc.edu}
\affiliation{Santa Cruz Institute for Particle Physics and Department of Physics, University of California Santa Cruz, 1156 High St., Santa Cruz, CA 95064, USA}

\author{P.~S.~Bhupal~Dev}
\email{bdev@wustl.edu}
\affiliation{Department of Physics and McDonnell Center for the Space Sciences,
Washington University, St.~Louis, MO 63130, USA}
\affiliation{PRISMA$^+$ Cluster of Excellence \& Mainz Institute for Theoretical Physics, 
Johannes Gutenberg-Universit\"{a}t Mainz, 55099 Mainz, Germany}

\author{Amarjit Soni\footnote{Senior physicist (Emeritus)}}
\email{adlersoni@gmail.com}
\affiliation{Physics Department, Brookhaven National Laboratory, Upton, NY 11973, USA}

\author{Fang~Xu}
\email{xufang@wustl.edu}
\affiliation{Department of Physics, Center for Field Theory and Particle Physics, Fudan University, Shanghai, 200433, China}
\affiliation{Department of Physics and McDonnell Center for the Space Sciences,
Washington University, St.~Louis, MO 63130, USA}

%%%%%%%%%%%%%%%%%%%%%%%%%%%%%%%
\begin{abstract}
{\bf Abstract:} We revisit the electric dipole moments (EDMs) of quarks and leptons in the Minimal Supersymmetric Standard Model (MSSM) with trilinear $R$-parity violation (RPV). In this framework, EDMs are induced at the two-loop level via RPV interactions. We perform a comprehensive recalculation of several classes of Barr--Zee type diagrams in a general $R_\xi$ gauge. While we find general agreement with previous analytic results in the literature, our work provides a valuable independent cross-check of the complicated calculations. We also point out some subtleties in the intermediate steps and in the choice of the flavor basis for the numerical evaluation of the expressions. By confronting the theoretical predictions with the latest experimental limits on EDMs, we derive updated constraints on combinations of RPV couplings. We highlight an approximate, testable correlation between the proton and neutron EDM that emerges within the considered class of RPV models, offering a distinctive signature for future EDM experiments.
\end{abstract}
%%%%%%%%%%%%%%%%%%%%%%%%%%%%%%%

\maketitle
\newpage
\tableofcontents
\newpage

%%%%%%%%%%%%%%%%%%%%%%%%%%%%%%%%%%%%%%%%%%%%%%%%%%%
\section{Introduction} \label{sec:intro}
%%%%%%%%%%%%%%%%%%%%%%%%%%%%%%%%%%%%%%%%%%%%%%%%%%%

The electric dipole moment (EDM) of a system is a $P$-odd and $T$-odd observable. Therefore, any observation of an EDM of an elementary particle would be a direct signal of $T$-violation, and hence, of $CP$-violation by virtue of the $CPT$-symmetry, required of any relativistic quantum field theory. Due to the smallness of Standard Model (SM) contributions to EDMs~\cite{Pospelov:1991zt, Pospelov:2013sca, Yamaguchi:2020eub, Yamaguchi:2020dsy, Ema:2022yra}, which are currently several orders of magnitude below the experimental sensitivities~\cite{Alarcon:2022ero}, EDMs provide a very clean probe of new sources of $CP$ violation (CPV) arising from beyond-the-SM (BSM) physics~\cite{Engel:2013lsa, Chupp:2017rkp, Alarcon:2022ero}. In fact, many BSM extensions with additional sources of CPV predict large EDMs within reach of experiments, and therefore, are severely constrained by existing experimental bounds on the neutron and electron EDMs.  

Based on dimensional analysis, new physics at a mass scale $\Lambda$, with a $CP$-violating phase $\delta$, contributing to the EDM of a fermion $f$ at the $n$-loop order will generically generate a fermion EDM of order
\begin{align}
    d_f \sim eq_f\left(\frac{g^2}{16\pi^2}\right)^n\xi_{\rm FV}\frac{m_f}{\Lambda^2}\sin{\delta} \, ,
\end{align}
where $m_f$ and $q_f$ are the mass and electric charge of the given fermion $f$, $g$ is the typical size of a coupling, and $\xi_{\rm FV}$ is a possible enhancement factor due to flavor violation. To give an example, consider the electron EDM: assume maximal $CP$-violation, i.e.~$\sin\delta=1$, $g\approx 0.6$ for $SU(2)_L$ weak gauge interactions, and $\xi_{\rm FV}=1$ for conservative estimates. Then the current experimental constraint of $|d_e|<4.8\times 10^{-30}e\,$cm~\cite{Roussy:2022cmp} translates into a lower bound on the new physics scale $\Lambda\gtrsim 70\, (3)$ TeV for $n=1\, (2)$. 
While an approximate estimate, this is more or less confirmed by concrete calculations in specific models, such as in the Two Higgs Doublet Model (2HDM)~\cite{Barr:1990vd, Abe:2013qla, Altmannshofer:2020shb, Altmannshofer:2024mbj}. This implies that current EDM searches are already sensitive to mass scales around or above those accessible at the LHC. Also, every improvement in the EDM bound by two orders of magnitude translates into an order of magnitude improvement in the mass reach. Given the fact that next-generation EDM experiments are expected to achieve a significant improvement in sensitivities (see Table~\ref{tab:EDM_exp}), it is worthwhile to consider well-motivated BSM scenarios that can be probed by such experiments. 

One such prime example of well-motivated BSM physics scenarios is the Minimal Supersymmetric Standard Model (MSSM)~\cite{Martin:1997ns}. It is arguably the best BSM candidate so far to address the gauge hierarchy problem,  origin of electroweak symmetry breaking and gauge coupling unification, among other things. Even assuming the minimal MSSM field content, the most general renormalizable, gauge-invariant superpotential allows terms that do not conserve $R$-parity. In fact, it has been argued~\cite{Brust:2011tb} that $R$-parity violation (RPV) is actually more natural than the ad hoc implementation of $R$-parity conservation in MSSM, usually performed to obtain a stable dark matter candidate. See Refs.~\cite{Altmannshofer:2017poe, Altmannshofer:2020axr,Dev:2021ipu, Afik:2022vpm} for other phenomenological consequences of this natural supersymmetry (SUSY) setup with RPV in various flavor and collider observables. 

There are two types of renormalizable RPV interaction terms, namely, bilinear and trilinear terms, allowed in the MSSM superpotential~\cite{Barbier:2004ez}. In general, they contain complex couplings, which are new sources of $CP$-violation, and hence, contribute to EDMs. It has been pointed out that in the absence of the bilinear terms, there are no additional fermion EDM contributions at the one-loop order~\cite{Godbole:1999ye, Abel:1999yz}, and the leading contributions come only at two-loop order~\cite{Chang:2000wf, Yamanaka:2012hm, Yamanaka:2012zq, Yamanaka:2012ep, Yamanaka:2012qn, Yamanaka:2013pfn}. Among these, the Barr--Zee type diagrams~\cite{Barr:1990vd} with photon exchange are typically the dominant ones, with subleading contributions from those with $W$ and $Z$ exchange. 

In this paper, we revisit the two-loop EDM calculations in the RPV-MSSM. Given the complexity of the calculations involved, this serves as a valuable independent check of the earlier calculations, which were mostly performed by a single author~\cite{Yamanaka:2012hm, Yamanaka:2012zq, Yamanaka:2012ep, Yamanaka:2012qn, Yamanaka:2013pfn}. 
While our final analytic expressions agree with those existing in the literature, we point out some subtleties in the intermediate steps and in the choice of the flavor basis for the numerical evaluation of the expressions. Finally, we present an updated summary of the current EDM constraints on the RPV couplings, taking into account the most recent experimental limits.   

We should clarify here that even without RPV, the MSSM has other possibilities to contain complex parameters that could become additional sources of CPV and could lead to sizable EDMs~\cite{Ellis:1982tk,  Buchmuller:1982ye, Polchinski:1983zd, delAguila:1983dfr, Dugan:1984qf, Nath:1991dn, Kizukuri:1992nj, Garisto:1996dj, Ibrahim:1997gj, Falk:1998pu, Ibrahim:1998je,  Chang:1998uc, Bartl:1999bc, Brhlik:1999ub, Pokorski:1999hz, Pilaftsis:1999td,  Barger:2001nu, Pilaftsis:2002fe, Demir:2003js, Altmannshofer:2008hc, Ellis:2008zy, Hisano:2008hn, Mercolli:2009ns, Paradisi:2009ey, Altmannshofer:2009ne, McKeen:2013dma, Altmannshofer:2013lfa, Cesarotti:2018huy}. Here,  we assume that the soft-breaking sector as well as the $\mu$-term in the superpotential are both $CP$-conserving. This choice of parameters is often suggested by supergravity~\cite{Chamseddine:1982jx, Dugan:1984qf, Arnowitt:1990eh,Accomando:1999uj}. In addition, we assume universality in the soft masses and the mixing of left- and right-handed squarks to avoid unwanted flavor-changing neutral currents 
at low energies -- the so-called minimal flavor violation hypothesis~\cite{DAmbrosio:2002vsn}. In our setup, the only source of $CP$-violation without RPV resides in the Yukawa couplings and can be described at the electroweak scale by the Cabibbo--Kobayashi--Maskawa (CKM) phase. In this case, the contribution from supersymmetric loops without RPV never exceeds the long-distance contribution to EDM from the usual SM~\cite{Gavela:1981sk, Khriplovich:1981ca,Romanino:1996cn,Hamzaoui:1997kt}. Therefore, we only consider here the EDMs sourced purely from RPV couplings.

The rest of the paper is structured as follows: In Section~\ref{sec:exp} we briefly review the current experimental status of EDM searches. In Section~\ref{sec:RPV} we describe the aspects of the RPV-MSSM that are most relevant to the discussion of EDMs. In Section~\ref{sec:EDMs}, we collect the most important two-loop contributions in our setup. Numerical results are presented in Section~\ref{sec:numerics}, updating and extending the results given in Refs.~\cite{Yamanaka:2012hm, Yamanaka:2012zq, Yamanaka:2012ep, Yamanaka:2012qn, Yamanaka:2013pfn}. We conclude in Section~\ref{sec:conclusions}. Technical details about some aspects of the two-loop calculation are given in Appendix~\ref{app:2loop}.

%%%%%%%%%%%%%%%%%%%%%%%%%%%%%%%%%%%%%%%%%%%%%%%%%%%
\section{Experimental status and prospects for EDM searches} \label{sec:exp}
%%%%%%%%%%%%%%%%%%%%%%%%%%%%%%%%%%%%%%%%%%%%%%%%%%%

The EDMs are $CP$-violating observables. So far, the only known source of CPV in the SM stems from the
CKM matrix in the quark sector. The resulting EDM diagrams necessarily involve at least two $W$-bosons
and multiple loops~\cite{Pospelov:1991zt, Pospelov:2013sca, Yamaguchi:2020eub, Yamaguchi:2020dsy, Ema:2022yra}, thus suppressing them many orders of magnitude below the current experimental sensitivities. Observation of an EDM at existing or planned experiments would be a clear sign of new physics. As of now, only upper limits on EDMs of elementary particles have been found.

EDMs of charged leptons, as well as EDMs and chromo EDMs (cEDMs) of quarks, are described by the following effective operators:
\begin{equation}
\mathcal L_\text{eff} =
 - \frac{d_f}{2} (\bar f \sigma^{\mu\nu} i\gamma_5 f) F_{\mu\nu}
 - \frac{g_s\tilde d_q}{2} (\bar q \sigma^{\mu\nu} i\gamma_5 T^a q) G^a_{\mu\nu} ~.
 \end{equation}
These operators feed into experimentally accessible $CP$-violating observables of composite systems, like frequency shifts in polar molecules, EDMs of nucleons and nuclei, or of diamagnetic atoms, see e.g. Refs.~\cite{Dekens:2018bci, Yamanaka:2022yoy, Kumar:2024yuu} for recent literature.

In the lepton sector, the by far strongest bounds have been obtained for the electron EDM. 
Recently, searches for the electron EDM using HfF$^+$ molecular ions~\cite{Roussy:2022cmp} have established an upper limit of\footnote{Note that Ref.~\cite{Roussy:2022cmp} reports a measurement of $d_e = (-1.3 \pm 2.0 \pm 0.6)\times 10^{-30} e\,\text{cm}$ and a limit of $|d_e| < 4.1 \times 10^{-30} e\,\text{cm}$ at 90\% confidence level (CL). Assuming a Gaussian likelihood, we translate this into the 95\% CL limit given in~\eqref{eq:de_exp}.} 
\begin{equation}\label{eq:de_exp}
 |d_e| < 4.8 \times 10^{-30} e\,\text{cm}  ~~~ @~95\% ~\text{CL}~,
\end{equation}
surpassing the previous best constraint from precision studies of ThO molecules by the ACME collaboration~\cite{ACME:2018yjb}. 
Improvements by an order of magnitude or more can be expected with future technologies~\cite{Alarcon:2022ero}. 
It is important to note that existing experimental limits on the electron EDM rely on measurements in paramagnetic systems such as HfF$^+$ ions or ThO molecules. While these systems are primarily sensitive to the electron EDM, their signals can also receive contributions from other sources of CPV, in particular $CP$-odd semileptonic interactions between electrons and nucleons. As a result, the interpretation of these measurements as constraints on the electron EDM requires some theoretical assumptions about the absence or suppression of such additional contributions;  see, for example, Ref.~\cite{Ardu:2025rqy} for a recent study. Here, we focus on the electron EDM contribution only.\footnote{In our considered RPV setup, four-fermion operators can arise that contribute to $CP$-odd electron-nucleon interactions. A careful analysis of such contributions goes beyond the scope of this paper, and we note that the numerical results that we obtain hold barring accidental cancellations.}

The existing constraints on the muon and tau EDMs are much weaker. The most stringent direct constraint on $d_\mu$ comes from the BNL muon $g-2$ experiment and reads \cite{Muong-2:2008ebm}
\begin{equation}
 |d_\mu| < 1.9 \times 10^{-19} e\,\text{cm} ~~~ @~95\% ~\text{CL}~.
\end{equation}
As noted in Refs.~\cite{Ema:2021jds, Ema:2022wxd}, a muon EDM can contribute at the loop level to EDMs of heavy atoms or molecules. Reinterpreting the ACME ThO result~\cite{ACME:2018yjb} in terms of a muon EDM gives an indirect bound $|d_\mu| < 1.7 \times 10^{-20} e$\,cm. It would be interesting to update the indirect constraint using the above-quoted HfF$^+$ result~\eqref{eq:de_exp}.

An improved direct sensitivity to the muon EDM down to $d_\mu \sim 10^{-21} e$\,cm can be expected from the Fermilab muon $g-2$ experiment~\cite{Chislett:2016jau}. Similarly, the planned muon $g-2$ experiment at JPARC is expected to search for a muon EDM with a sensitivity goal of $d_\mu \sim 1.5 \times 10^{-21} e$\,cm~\cite{Abe:2019thb}. This means both the Fermilab and the JPARC experiments can achieve an improvement over the existing direct bound by 2 orders of magnitude.
Moreover, there are ideas to probe the muon EDM with a precision of $d_\mu \sim 6 \times 10^{-23} e$\,cm and potentially even further at PSI~\cite{Adelmann:2010zz, Adelmann:2021udj}.

The current most stringent constraint on the tau EDM is obtained from precision measurements of $e^+ e^- \to \tau^+ \tau^-$ at Belle~\cite{Belle:2021ybo}:
\begin{eqnarray}
  -1.85 \times 10^{-17} e\,\text{cm} < &\text{Re}(d_\tau)& < 0.61 \times 10^{-17} e\,\text{cm}   ~~~ @~95\% ~\text{CL}~, \\
  -1.03 \times 10^{-17} e\,\text{cm}  < &\text{Im}(d_\tau)& < 0.23 \times 10^{-17} e\,\text{cm}  ~~~ @~95\% ~\text{CL}~.
\end{eqnarray}
Note that the Belle measurements do not probe the tau EDM at zero momentum transfer, but at $\sqrt{s} = 10.58$\,GeV. For a non-zero momentum transfer, the tau EDM can be a complex quantity, and real and imaginary parts can be constrained separately.
Belle II is expected to reach sensitivities in the ballpark of $10^{-18}e$\,cm to $10^{-19}e$\,cm~\cite{Belle-II:2018jsg}.
A Belle II upgrade with beam polarization might improve the sensitivity further to a level of $\sim 10^{-20}e$\,cm~\cite{USBelleIIGroup:2022qro}. 

Analogous to the muon EDM, there are also indirect constraints on the tau EDM due to its higher-order contributions to EDMs of heavy atoms and molecules. Using the ACME limit from Ref.~\cite{ACME:2018yjb}, Ref.~\cite{Ema:2022wxd} finds $|\text{Re}(d_\tau)| < 1.1 \times 10^{-18} e$\,cm (see also Refs.~\cite{Grozin:2008nw, Ema:2021jds} for earlier studies).

Constraints on the tau EDM can also be obtained at high-energy colliders. Ref.~\cite{Haisch:2023upo} estimates a constraint of $|\text{Re}(d_\tau)| < 1.0 \times 10^{-17} e$\,cm based on $pp \to \tau^+ \tau^-$ results at momentum transfers of $\mathcal O(1\, \text{TeV})$. Using ultra peripheral Pb-Pb collisions at the LHC might be another avenue to obtain constraints on the tau EDM~\cite{Baltz:2007kq}, as demonstrated by the recent ATLAS and CMS measurements of the tau magnetic dipole moment~\cite{ATLAS:2022ryk,CMS:2022arf}. Similarly, the highly linear polarization of coherent photons in $\gamma\gamma\to \tau^+\tau^-$ at future lepton colliders can also constrain the tau EDM to $|\text{Re}(d_\tau)| < 2.8 \times 10^{-16} e$\,cm~\cite{Shao:2025xwp}.

The fact that the different limits on the tau EDM are obtained at different values of momentum transfer is relevant for the interpretation in terms of light new physics, but does not play a role in constraining heavy new physics. The contributions to the tau EDM that we discuss in this work arise from new physics that is much heavier than the typical momentum transfer. Such contributions are real and are therefore subject to the constraints on $\text{Re}(d_\tau)$.

\bigskip
Quark EDMs and cEDMs feed into EDMs of hadronic systems, in particular the neutron and proton, but also heavy nuclei. In the following, we focus on the neutron and proton and use the approximate expressions given in Ref.~\cite{Kaneta:2023wrl}:
\begin{eqnarray} \label{eq:dn}
d_n &\simeq& 0.73\, d_d - 0.18\, d_u + e (0.20\, \tilde d_d + 0.10\, \tilde d_u)~, \\ \label{eq:dp}
d_p &\simeq& 0.73\, d_u - 0.18\, d_d - e (0.40\, \tilde d_u + 0.049\, \tilde d_d)~.
\end{eqnarray}
The numerical coefficients that relate the quark EDMs to the neutron and proton EDMs are known with fairly high precision, in particular from lattice calculations~\cite{Yamanaka:2018uud, Gupta:2018qil, Gupta:2018lvp, Alexandrou:2019brg, Park:2020axe, FlavourLatticeAveragingGroupFLAG:2024oxs} that have achieved better than 10\% precision.
The corresponding coefficients of the quark cEDMs are subject to sizable, $\mathcal O(1)$, uncertainties. The values quoted above have been obtained in Ref.~\cite{Kaneta:2023wrl} and are based on a combination QCD sum rule and lattice results~\cite{Belyaev:1982sa, Ball:2002ps, Bali:2012jv}.
In our numerical analysis in Section~\ref{sec:numerics}, we will simply use the above central values, neglecting the theoretical uncertainties.
It illustrates what could be achieved with present data for future
improved theory calculations of the hadronic, nuclear, and atomic matrix elements.
Additional contributions from the strange quark EDM and cEDM have very large uncertainties.
Lattice calculations indicate that the strange quark contribution is very small~\cite{Bhattacharya:2016zcn, Gupta:2018lvp, Park:2025rxi}, while~\cite{Vecchi:2025jbb} argues that the strange quark contribution might be relevant.
We will neglect the strange contribution in our numerical analysis.
Additional contributions to the neutron and proton EDMs can come from the Weinberg three-gluon operator and from $CP$-odd four-quark operators (see e.g. Refs.~\cite{Demir:2002gg, deVries:2012ab, Dekens:2018bci, Yamanaka:2022qlu, Osamura:2022rak}). Such contributions come with sizable hadronic uncertainties, and a comprehensive analysis is beyond the scope of this work. We will neglect them in the following and focus instead on the quark dipole moments.

On the experimental side, the current best constraint on the neutron EDM has been obtained at the PSI using ultracold neutrons~\cite{Abel:2020pzs}:\footnote{Note that Ref.~\cite{Abel:2020pzs} reports a measurement of $d_n = (0.0 \pm 1.1 \pm 0.2)\times 10^{-26} e\,\text{cm}$ and a limit of $|d_n| < 1.8 \times 10^{-26} e\,\text{cm}$ at 90\% CL. Assuming a Gaussian likelihood, we translate this into the 95\% CL limit given in~\eqref{eq:dn_exp}.}
\begin{equation} \label{eq:dn_exp}
   |d_n| < 2.2 \times 10^{-26} e\,\text{cm}  ~~~ @~95\% ~\text{CL}~.
\end{equation}
Ongoing work aims to improve the sensitivity to the level of $\sim 10^{-27} e$\,cm and beyond~\cite{nEDM:2019qgk, Wurm:2019yfj, n2EDM:2021yah}.

No direct limit on the proton EDM has been obtained so far. The best indirect limit is obtained by reinterpreting a search for the dipole moment of ${}^{199}$Hg~\cite{Graner:2016ses} and is given by~\cite{Chupp:2017rkp}
\begin{equation}
|d_p| < 2.0 \times 10^{-25} e\,\text{cm} ~~~ @~95\% ~\text{CL}~.
\end{equation}
This limit is inferred from the ${}^{199}$Hg atomic EDM through its sensitivity to the short-range proton contribution. Other $CP$-violating sources, such as those arising from quark chromo-EDMs and the induced $CP$-odd pion–nucleon couplings, may also contribute to the atomic EDM via hadronic and nuclear effects. Although in the framework considered here, the chromo-EDM amplitudes are typically smaller than the photon- and electroweak-induced quark EDMs, implying that such additional contributions are expected to be moderate, the use of the ${}^{199}$Hg-based bound still carries some degree of model dependence and theoretical uncertainty.
Proposals for a storage ring experiment searching \textcolor{red}{directly} for the proton EDM might reach sensitivities down to $\sim 10^{-29} e$\,cm~\cite{Farley:2003wt, Anastassopoulos:2015ura, CPEDM:2019nwp, Alexander:2022rmq}.

The current bounds and expected sensitivities to the various EDMs are summarized in Table~\ref{tab:EDM_exp}. These values will be contrasted below with our theoretical predictions in $R$-parity violating MSSM. 

\renewcommand{\arraystretch}{1.5}
%%%%%%%%%%%%%%%%%%%%%%%%%%%%%%%%%
\begin{table}[tb]
\begin{ruledtabular}
\begin{tabular}{cccc}
observable & direct limit & indirect limit & projected sensitivity \\
\hline
$|d_e|$ & $< 4.8 \times 10^{-30} \,e\,\text{cm}$ \cite{Roussy:2022cmp} & & $ \sim 10^{-30} \,e\,\text{cm}$ \cite{ACME:2018yjb}\\ \hline
$|d_\mu|$ & $ < 1.9 \times 10^{-19} \,e\,\text{cm}$ \cite{Muong-2:2008ebm} & $< 1.7 \times 10^{-20} \,e\,\text{cm}$ \cite{ACME:2018yjb, Ema:2021jds, Ema:2022wxd} & $ \sim 10^{-21} \,e\,\text{cm}$ \cite{Chislett:2016jau,Abe:2019thb} \\
 & & & $ \sim 6 \times 10^{-23} \,e\,\text{cm}$ \cite{Adelmann:2010zz,Adelmann:2021udj} \\ \hline
$|\text{Re}(d_\tau)|$ & $ < 1.7 \times 10^{-17} \,e\,\text{cm}$ \cite{Belle:2021ybo} & $< 1.1 \times 10^{-18} \,e\,\text{cm}$ \cite{ACME:2018yjb, Ema:2021jds, Ema:2022wxd} & $ \sim 10^{-19} \,e\,\text{cm}$ \cite{Belle-II:2018jsg} \\ 
 & & & $ \sim 10^{-20} \,e\,\text{cm}$ \cite{USBelleIIGroup:2022qro} \\
\hline\hline
$|d_n|$ & $ < 2.2 \times 10^{-26} \,e\,\text{cm}$ \cite{Abel:2020pzs} & & $ 10^{-27} - 10^{-28} \,e\,\text{cm}$ \cite{nEDM:2019qgk, Wurm:2019yfj, n2EDM:2021yah}  \\
\hline
$|d_p|$ & --- & $< 2.0 \times 10^{-25} e\,\text{cm}$ \cite{Graner:2016ses, Chupp:2017rkp} & $\sim 10^{-29} \,e\,\text{cm}$ ~\cite{Farley:2003wt, Anastassopoulos:2015ura, CPEDM:2019nwp, Alexander:2022rmq} \\
\end{tabular}
\end{ruledtabular}
\caption{Summary of current experimental limits and projected precision of next-generation experiments searching for EDMs of the charged leptons, the neutron, and the proton. The indirect limits of the muon and tau EDMs are based on loop contributions to atomic and molecular EDMs. The indirect limit on the proton EDM is coming from the constraint on the EDM of ${}^{199}$Hg.}
\label{tab:EDM_exp}
\end{table}
%%%%%%%%%%%%%%%%%%%%%%%%%%%%%%%%%
\renewcommand{\arraystretch}{1.}

%%%%%%%%%%%%%%%%%%%%%%%%%%%%%%%%%%%%%%%%%%%%%%%%%%%
\section{The MSSM with $R$-parity Violation} \label{sec:RPV}
%%%%%%%%%%%%%%%%%%%%%%%%%%%%%%%%%%%%%%%%%%%%%%%%%%%

Our setup is the MSSM with $R$-parity violating interactions; see Ref.~\cite{Barbier:2004ez} for a review.
Among the renormalizable $R$-parity violating couplings, we consider the trilinear $LLE$ and $LQD$ terms that violate lepton number but preserve baryon number. We do not consider the baryon number violating $UDD$ terms, as they do not contribute to the Barr--Zee type two-loop contributions to EDMs that we focus on in this paper. In the presence of both lepton number and baryon number violating interactions, one would obtain extremely strong bounds from proton decay~\cite{Nath:2006ut}. We also do not consider bilinear $R$-parity violating terms. Such terms lead to Higgs-slepton mixing, chargino-lepton mixing, and neutralino-neutrino mixing, which are very tightly constrained~\cite{Barbier:2004ez}.
The superpotential we work with is therefore given by
\begin{equation} \label{eq:W_RPV}
    W_\text{RPV} = \frac{1}{2} \lambda_{ijk} \hat L_i \hat L_j \hat E^c_k + \lambda^\prime_{ijk} \hat L_i \hat Q_j \hat D^c_k ~.
\end{equation}
The superfields $\hat L$, $\hat E^c$, $\hat Q$, $\hat D^c$ correspond to the left-handed lepton doublets, the right-handed lepton singlets, the left-handed quark doublets, and the right-handed down quark singlets, respectively. The $i,j,k$ are flavor indices. Note that the $LLE$ couplings are antisymmetric in the first two flavor indices $\lambda_{ijk} = - \lambda_{jik}$. 

The above superpotential gives the following lepton-number violating interactions of fermions with sfermions:
\begin{eqnarray} \label{eq:LLE}
    \mathcal L_{LLE} &\supset& - \frac{1}{2} \lambda_{ijk} \Big( \tilde \nu_i \bar e_{R_k} e_{L_j} + \tilde e_{L_j} \bar e_{R_k} \nu_i + \tilde e_{R_k}^* \bar \nu_i^c e_{L_j} - (i \leftrightarrow j) \Big) + ~\text{H.c.} ~,  \\[12pt]
 \label{eq:LQD}    \mathcal L_{LQD} &\supset& - \lambda^\prime_{ijk} \Big( \tilde \nu_i \bar d_{R_k} d_{L_j} + \tilde d_{L_j} \bar d_{R_k} \nu_i + \tilde d^*_{R_k} \bar \nu_i^c d_{L_j} \Big)   + ~\text{H.c.} \nonumber \\ 
    && \qquad \qquad \qquad +  \lambda^\prime_{imk} V^*_{jm} \Big( \tilde e_{L_i} \bar d_{R_k} u_{L_j} + \tilde u_{L_j} \bar d_{R_k} e_{L_i} + \tilde d^*_{R_k} \bar e_{R_i}^c u_{L_j} \Big) + ~\text{H.c.} ~.
\end{eqnarray}
Here, $e_L$, $e_R$, $d_L$, $d_R$ and $u_L$ are left-handed and right-handed charged leptons and quarks, $\nu$ are the left-handed neutrinos and $\tilde e_L$, $\tilde e_R$, $\tilde d_L$, $\tilde d_R$, $\tilde u_L$, $\tilde \nu$ are the corresponding spin-0 superpartners. In writing Eqs.~\eqref{eq:LLE} and~\eqref{eq:LQD}, we work in a flavor basis in which all charged fermions correspond to mass eigenstates with the corresponding sfermions rotated in the same way (this is sometimes referred to as the ``super CKM basis''). We define the $\lambda^\prime$ couplings such that they directly correspond to the RPV interactions of left-handed down quarks and squarks. The corresponding couplings of the left-handed up quark and squarks pick up a CKM factor $V_{jm}^*$ in this convention.
We neglect neutrino mixing as it is of no relevance for our calculations presented here.

The superpotential terms in Eq.~\eqref{eq:W_RPV} also give rise to trilinear $R$-parity violating interactions of scalars that enter the two-loop EDMs. In particular,  when combining the $LLE$ and $LQD$ interactions with the Yukawa couplings, one gets the following $F$ terms
\begin{eqnarray}
    \mathcal L_{LLE} &~\to~& - \lambda_{ijk} \Big( m_{e_j} \tilde \nu_i ~\tilde e_{R_k}^* \tilde e_{R_j} + m_{e_k} \tilde \nu_i ~ \tilde e_{L_k}^* \tilde e_{L_j} \Big) ~ + ~\text{H.c.} ~,  \label{eq:scalar int1}\\[12pt]
    \mathcal L_{LQD} &~\to~& - \lambda^\prime_{ijk} \Big( m_{d_j} \tilde \nu_i~ \tilde d_{R_k}^* \tilde d_{R_j} + m_{d_k} \tilde \nu_i~ \tilde d_{L_k}^* \tilde d_{L_j} \Big) + \lambda^\prime_{imk} V^*_{jm} m_{d_k} \tilde e_{L_i} ~ \tilde d_{L_k}^* \tilde u_{L_j}  ~ + ~\text{H.c.} ~. \label{eq:scalar int2}
\end{eqnarray}
Furthermore, there are also the corresponding $R$-parity violating soft SUSY-breaking terms
\begin{eqnarray}
    \mathcal L_{LLE}^\text{soft} & = & \frac{1}{2} A_{ijk} \Big( \tilde \nu_i ~ \tilde e_{R_k}^* \tilde e_{L_j} - (i \leftrightarrow j) \Big)  ~+~\text{H.c.} ~, \\[12pt]
    \mathcal L_{LQD}^\text{soft} & = & A^\prime_{ijk} \tilde \nu_i ~ \tilde e_{R_k}^* \tilde d_{L_j} -  A^\prime_{imk} V^*_{jm} \tilde e_{L_i} \tilde d_{R_k}^* \tilde u_{L_j}   ~ + ~\text{H.c.} ~.
\end{eqnarray}
Analogous to $\lambda_{ijk}$, $A_{ijk}$ is also antisymmetric in its first two indices $A_{ijk} = - A_{jik}$. 
Such terms are typically not considered in the literature and are likewise omitted in the present analysis. They can enter the predictions of the EDMs at the two-loop level. Studying their effects is left for future work.

%\bigskip
Before we start discussing the RPV contributions to EDMs, a comment about the particle spectrum and the assumed hierarchy in masses is in order. 

The bulk of the masses of the SUSY particles is due to the usual $R$-parity conserving soft-SUSY breaking terms. In this paper, we are interested in the genuine effect of RPV on EDMs. We thus assume that none of the $R$-parity conserving SUSY breaking terms (and the Higgsino mass) introduce any $CP$-violating phases. In addition, we will for simplicity, assume that all sfermion masses and $R$-parity conserving trilinear couplings are flavor diagonal in the fermion mass basis. 

Given the absence of any direct sign of SUSY particles at the LHC, it is a reasonable assumption that there is a significant mass gap between the SUSY scale and the weak scale, $m_\text{SUSY} \gg v$. Furthermore, the spectrum of the SM particles is hierarchical, with only a handful of particles, namely, the top quark, the $W$ and $Z$ bosons, and the Higgs boson, close to the weak scale, while all other particles are much lighter. Whenever it is appropriate, we will expand our results for the EDMs in small mass ratios and only keep the leading terms.

The hierarchy $m_\text{SUSY} \gg v$ also implies that left-right mixing of sfermions can usually be neglected or treated as perturbation if necessary.

%%%%%%%%%%%%%%%%%%%%%%%%%%%%%%%%%%%%%%%%%%%%%%%%%%%
\section{RPV Contributions to EDMs}  \label{sec:EDMs}
%%%%%%%%%%%%%%%%%%%%%%%%%%%%%%%%%%%%%%%%%%%%%%%%%%%

%%%%%%%%%%%%%%%%%%%%%%%%%%%%%%%%%%%%%%
\subsection{One-loop contributions}
%%%%%%%%%%%%%%%%%%%%%%%%%%%%%%%%%%%%%%

Despite the large number of $R$-parity violating couplings in the superpotential, none of them give new contributions to electric and chromo-electric dipole moments of charged leptons and quarks at the one-loop level. While magnetic dipole moments do receive contributions at one-loop, it has been shown that new contributions to EDMs and cEDMs arise first at two-loop~\cite{Godbole:1999ye, Abel:1999yz, Chang:2000wf}. This is due to the chirality structure of the RPV couplings, which is such that the same scalar does not couple to both left-handed and right-handed versions of a given fermion. As a consequence, and using power counting rules for loop diagrams, one finds that 1-loop corrections to fermion dipole moments are necessarily proportional to the absolute value squared of trilinear RPV couplings and therefore do not provide a $CP$-violating phase.

%%%%%%%%%%%%%%%%%%%%%%%%%%%%%%%%%%%%%%
\subsection{Two-loop contributions}
\begin{figure}[htb]
\centering
\includegraphics[width=0.32\linewidth]{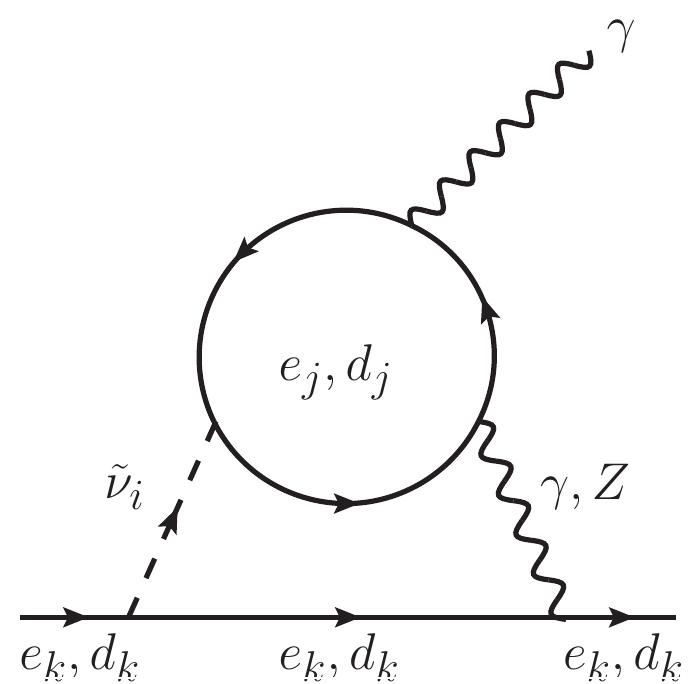}
\includegraphics[width=0.32\linewidth]{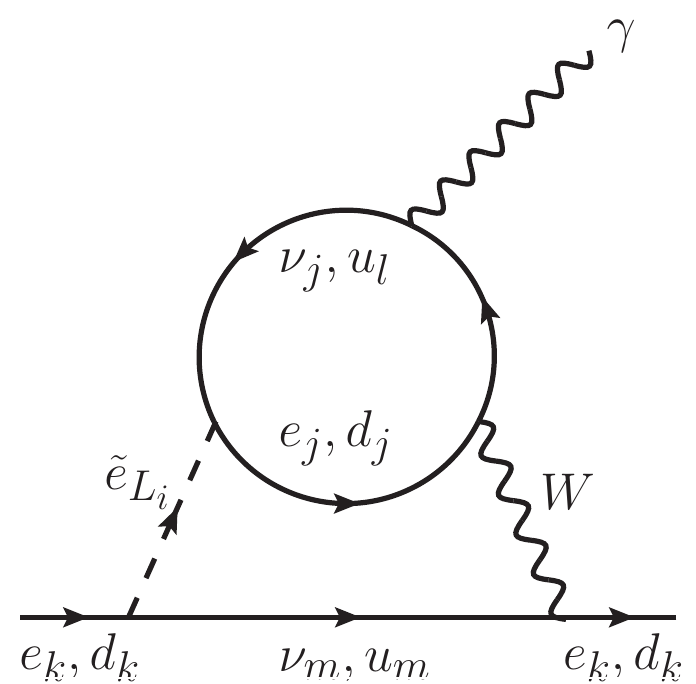}
\includegraphics[width=0.32\linewidth]{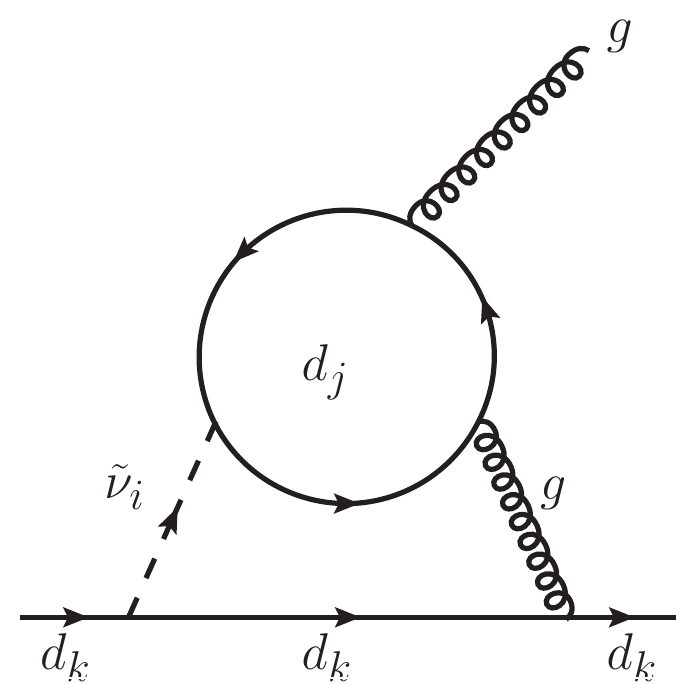}
\caption{Example Barr--Zee diagrams with lepton or quark loops, that contribute to lepton and quark EDMs (left and center) or quark chromo EDMs (right).}
\label{fig:BarrZee_fermion}
\end{figure}

\begin{figure}[tbh]
\centering
\includegraphics[width=0.32\linewidth]{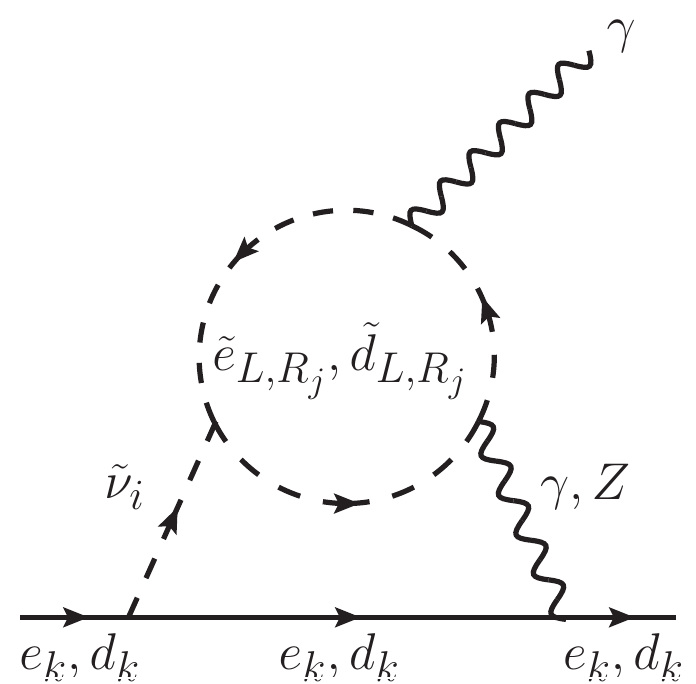}
\includegraphics[width=0.32\linewidth]{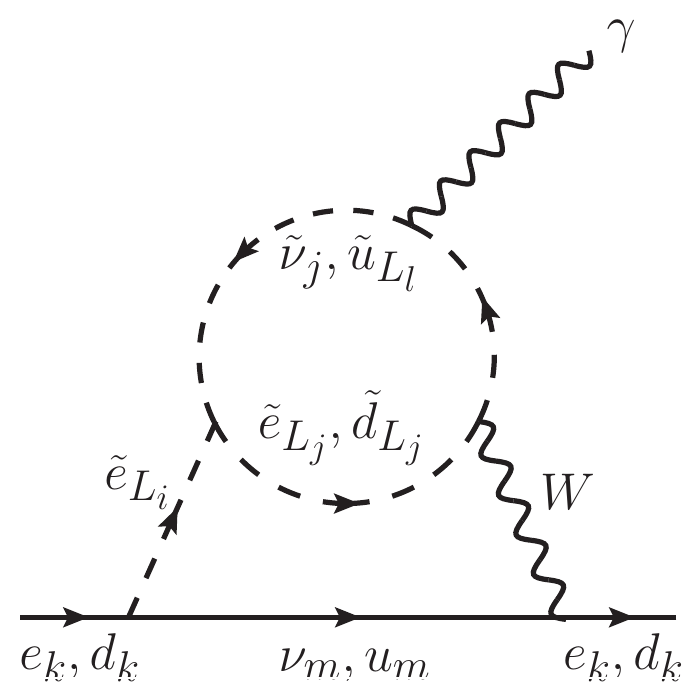}
\includegraphics[width=0.32\linewidth]{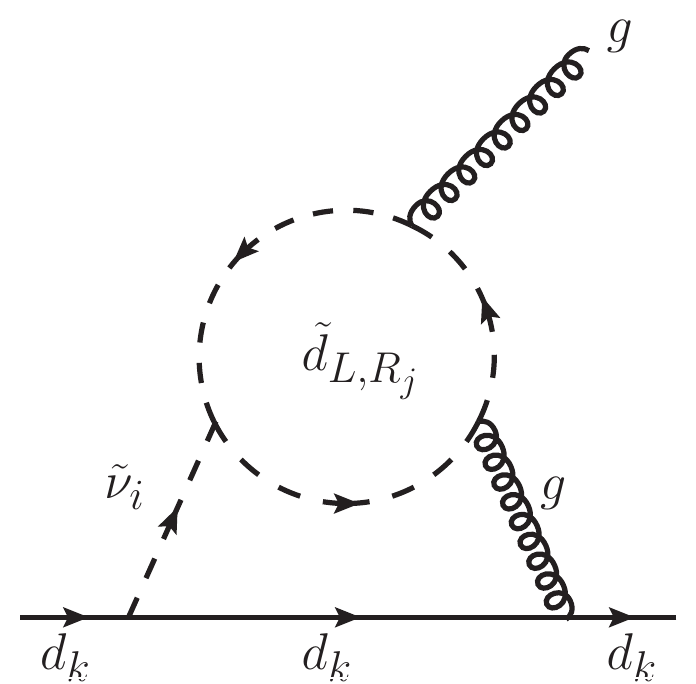}
\caption{Example Barr--Zee diagrams with slepton or squark loops, that contribute to lepton and quark EDMs (left and center) or quark chromo EDMs (right).}
\label{fig:BarrZee_sfermion}
\end{figure}

\begin{figure}[htb]
\centering
\includegraphics[width=0.32\linewidth]{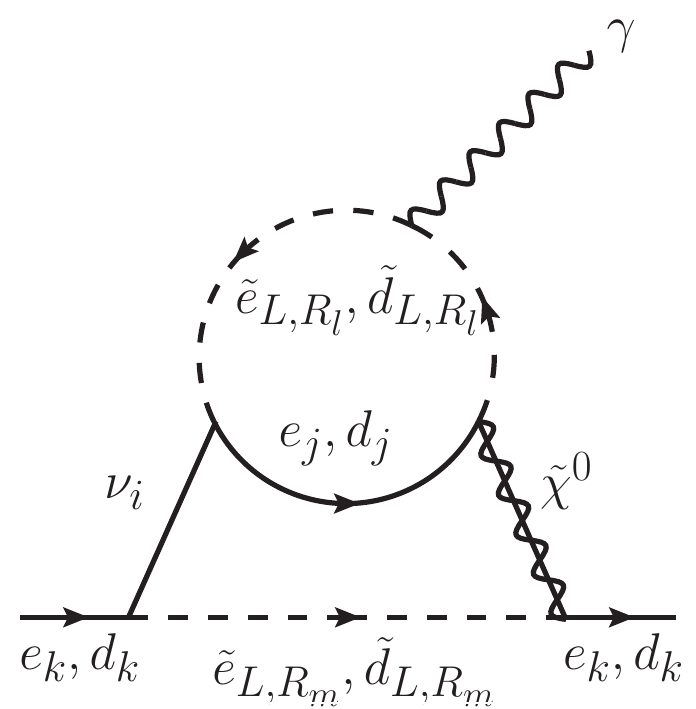}
\includegraphics[width=0.32\linewidth]{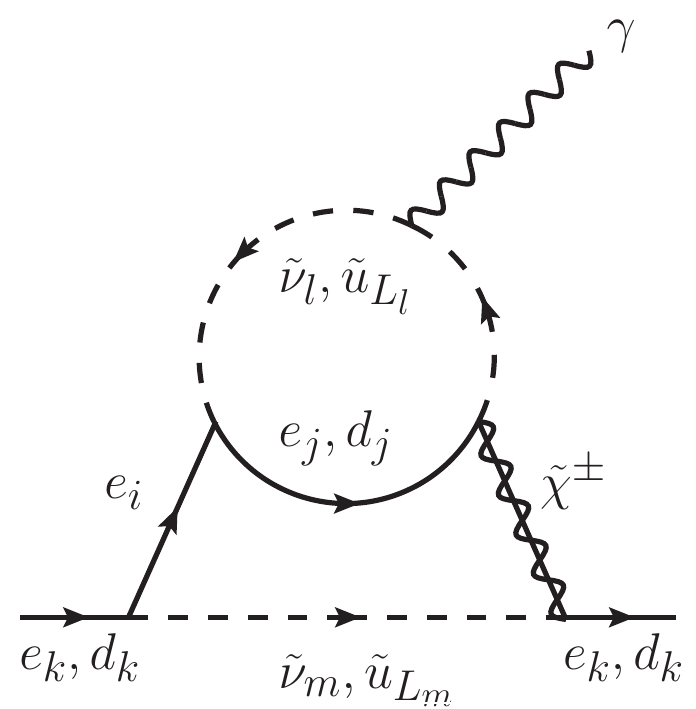}
\includegraphics[width=0.32\linewidth]{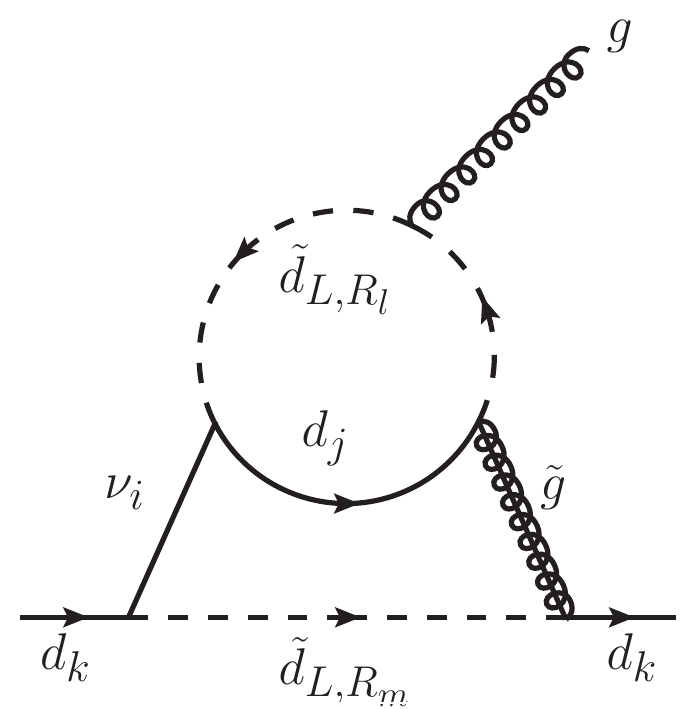}
\caption{Example two-loop diagrams involving neutralinos, charginos, and gluinos that contribute to lepton and quark EDMs (left and center) or quark chromo EDMs (right). In the first and third panels, we do not indicate the arrows on the neutrino fermion lines, as their directions are reversed depending on whether the left- or right-handed sfermion is considered.}
\label{fig:gauginos}
\end{figure}

%%%%%%%%%%%%%%%%%%%%%%%%%%%%%%%%%%%%%%
There are various classes of two-loop contributions to lepton and quark EDMs in the MSSM with RPV. Fig.~\ref{fig:BarrZee_fermion} shows example Barr--Zee type diagrams~\cite{Barr:1990vd} that include a loop of leptons or quarks as well as SM gauge bosons. Analytical expression for those contributions have been calculated in Refs.~\cite{Yamanaka:2012hm,Yamanaka:2012ep}. Instead of leptons or quarks, the loop can also contain sleptons or squarks. Example diagrams of this type are shown in Fig.~\ref{fig:BarrZee_sfermion}. These contributions have been evaluated in Refs.~\cite{Yamanaka:2012zq,Yamanaka:2012ep}. There are also related diagrams that, instead of gauge bosons, contain neutralinos, charginos, or gluinos. Examples are given in Fig.~\ref{fig:gauginos}. Expressions for these type of diagrams have been obtained in Ref.~\cite{Yamanaka:2012qn}. 
Various additional two-loop contributions that involve two RPV couplings but that are not of the Barr--Zee type have been identified in Ref.~\cite{Chang:2000wf}. However, as discussed in Ref.~\cite{Chang:2000wf}, such contributions necessarily involve charged Higgs bosons and are suppressed by light quark Yukawa couplings and CKM mixing angles. We will not discuss these latter contributions in our work.
%%%%%%%%%%%%%%%%%%%%%%%%%%%%%%%%%%%%%%
%%%%%%%%%%%%%%%%%%%%%%%%%%%%%%%%%%%%%%
%%%%%%%%%%%%%%%%%%%%%%%%%%%%%%%%%%%%%%

The contributions mentioned so far all contain exactly two RPV couplings. There are additional two-loop contributions in RPV SUSY that contain four RPV couplings. Order of magnitude estimates for some of these contributions have been discussed in Ref.~\cite{Abel:1999yz}. To the best of our knowledge, no explicit calculation of those diagrams exists in the literature. 

The total two-loop RPV contributions to the EDM of a lepton or a down-type quark $d_{\ell,d}$ and chromo EDM of a down-type quark $\tilde{d_d}$ can thus be written as
\begin{eqnarray}
    d_\ell^\text{RPV} &=&  d_\ell^{f,\gamma} + d_\ell^{f,Z} + d_\ell^{f,W} + d_\ell^{\tilde f,\gamma} + d_\ell^{\tilde f,Z} + d_\ell^{\tilde f,W} + d_\ell^{\tilde \chi^0} +d_\ell^{\tilde \chi^\pm} + d_\ell^{4\lambda} ~, \\
    d_d^\text{RPV} &=&  d_d^{f,\gamma} + d_d^{f,Z} + d_d^{f,W} + d_d^{\tilde f,\gamma} + d_d^{\tilde f,Z} + d_d^{\tilde f,W} + d_d^{\tilde \chi^0} +d_d^{\tilde \chi^\pm} + d_d^{4\lambda} ~, \\
    \tilde d_d^\text{RPV} &=&  \tilde d_d^{f,g} + \tilde d_d^{\tilde f, g} + \tilde d_d^{\tilde g} + \tilde d_d^{4\lambda} ~.
\end{eqnarray}
Note that the classes of diagrams we have mentioned here do not give two-loop RPV contributions to up-type EDMs or chromo-EDMs. As discussed in Refs.~\cite{Godbole:1999ye, Chang:2000wf}, there are two-loop contributions to up-type EDMs and chromo-EDMs proportional to $\lambda^\prime$ couplings that feature charged Higgs bosons in the loops. These contributions are suppressed by small Yukawa couplings or small CKM matrix elements, and we do not consider them here. Two-loop contributions to up-type EDMs and chromo-EDMs can also arise in the presence of $UDD$-type RPV couplings~\cite{Barbieri:1985ty}, which are however severely suppressed by the proton decay constraints. 
This restriction has interesting implications for the proton EDM as we will discuss in Section~\ref{sec:proton}.

The goal of our paper is to carefully re-evaluate the bulk of the RPV contributions to the EDMs and chromo-EDMs, namely the Barr--Zee contributions with SM gauge bosons and fermion or sfermion loops, and to update the corresponding constraints on the RPV couplings from existing bounds on EDMs. 
A detailed discussion of the remaining two-loop contributions, namely those that involve gauginos and the ones that depend on four powers of RPV couplings, will be presented elsewhere.

We have performed our calculation in a generic $R_\xi$ background field gauge and have carefully monitored the effect of $\gamma_5$, making use of the \textquotesingle t~Hooft--Veltman scheme. Some technical details are given in appendix~\ref{app:2loop}.
In many of our results, the Davydychev--Tausk function appears~\cite{Davydychev:1992mt}:
\begin{equation}
\Phi(x) = \frac{1}{\sqrt{1-4x}} \left[ \frac{\pi^2}{3} - \log^2(x) + 2 \log^2\Big(\frac{1- \sqrt{1-4x}}{2}\Big) - 4 {\rm Li}_2\Big(\frac{1- \sqrt{1-4x}}{2} \Big) \right] ~,
\end{equation}
where ${\rm Li}_2(x) = - \int_0^x {\rm d}z \,\log(1-z)/z$ is the di-logarithm function.
We find that the Davydychev--Tausk function is related to a Feynman parameter integral that we encountered frequently in our calculations
\begin{equation}
\int_0^1 dz ~ \frac{\log\big( \frac{x}{z(1-z)} \big)}{1 - \frac{x}{z(1-z)}}= 2 + \log(x) - x \Phi(x) ~.
\end{equation}
The first terms in the small $x$ expansion of the Davydychev--Tausk function are
\begin{equation}
    \Phi(x) \simeq \log^2(x) + \frac{\pi^2}{3} + ... ~.
\end{equation}

\bigskip
%%%%%%%%%%%%%%%%%%%%%%%%%%%%%%%%%%%%%%%%%%%%%%%%
\paragraph{Barr--Zee diagrams with SM fermions:} We have recalculated the Barr--Zee diagrams with fermions (c.f. Fig.~\ref{fig:BarrZee_fermion}) and find for a lepton $e_k$, in agreement with Refs.~\cite{Yamanaka:2012hm, Yamanaka:2012ep}, the following contributions from photon diagrams, $Z$ boson diagrams and $W$ boson diagrams: 
\begin{align}
 \frac{d_{e_k}^{f,\gamma}}{e} = \frac{\alpha_{\rm em}}{16 \pi^3} \sum_{i,j} \Bigg[ &\text{Im}\big(\lambda_{ijj} \lambda_{ikk}^* \big) Q^2_{e_j}Q_{e_k}  \frac{m_{e_j}}{m_{\tilde \nu_{i}}^2} \bigg( \log\big( \frac{m_{\tilde \nu_{i}}^2}{m_{e_j}^2} \big) - 2 + \frac{m_{e_j}^2}{m_{\tilde \nu_{i}}^2} \Phi\big( \frac{m_{e_j}^2}{m_{\tilde \nu_{i}}^2} \big)\bigg) \notag\\ 
 + &\text{Im}\big(\lambda_{ijj}^\prime \lambda_{ikk}^{*} \big) n_c Q^2_{d_j}Q_{e_k}  \frac{m_{d_j}}{m_{\tilde \nu_{i}}^2} \bigg( \log\big( \frac{m_{\tilde \nu_{i}}^2}{m_{d_j}^2} \big) - 2 + \frac{m_{d_j}^2}{m_{\tilde \nu_{i}}^2} \Phi\big( \frac{m_{d_j}^2}{m_{\tilde \nu_{i}}^2} \big)\bigg)  \Bigg] ~, 
\label{eq:lk f gamma}
\end{align}
\begin{align}
 \frac{d_{e_k}^{f,Z}}{e} = &\frac{\alpha_{\rm em}}{16 \pi^3} \sum_{i,j}  \frac{1}{1 - m_Z^2/m_{\tilde \nu_{i}}^2}  \Bigg[ \notag\\
 &\text{Im}\big(\lambda_{ijj} \lambda_{ikk}^* \big) Q_{e_j}\alpha_{e_j}\alpha_{e_k} \frac{m_{e_j}}{m_{\tilde \nu_{i}}^2} \bigg( \log\big( \frac{m_{\tilde \nu_{i}}^2}{m_Z^2} \big) - \frac{m_{e_j}^2}{m_Z^2} \Phi\big(\frac{m_{e_j}^2}{m_Z^2}\big) + \frac{m_{e_j}^2}{m_{\tilde \nu_{i}}^2} \Phi\big(\frac{m_{e_j}^2}{m_{\tilde \nu_{i}}^2}\big)\bigg) \notag\\
 + &\text{Im}\big(\lambda_{ijj}^\prime \lambda_{ikk}^{*} \big) n_c Q_{d_j}\alpha_{d_j}\alpha_{e_k} \frac{m_{d_j}}{m_{\tilde \nu_{i}}^2} \bigg( \log\big( \frac{m_{\tilde \nu_{i}}^2}{m_Z^2} \big) - \frac{m_{d_j}^2}{m_Z^2} \Phi\big(\frac{m_{d_j}^2}{m_Z^2}\big) + \frac{m_{d_j}^2}{m_{\tilde \nu_{i}}^2} \Phi\big(\frac{m_{d_j}^2}{m_{\tilde \nu_{i}}^2}\big)\bigg)   \Bigg]   ~,
\label{eq:lk f Z}
\end{align}
\begin{align}
 \frac{d_{e_k}^{f,W}}{e} =& -\frac{\alpha_{\rm em}}{128 \pi^3 s_W^2} \int_0^1 dz~ \Bigg[ \sum_{i,j} \text{Im}\big(\lambda_{ijj} \lambda_{ikk}^* \big) Q_{e_j} z \frac{m_{e_j}}{m_{\tilde e_{L_{i}}}^2} \tilde{C}_0\Big( 1 , \frac{m_W^2}{m_{\tilde e_{L_{i}}}^2} , \frac{m_{e_j}^2}{(1-z) m_{\tilde e_{L_{i}}}^2} \Big) \notag\\ 
 + \sum_{i,j,l,n}  &\text{Im}\big(\lambda_{inj}^\prime \lambda_{ikk}^{*} V_{lj} V_{ln}^* \big) n_c\big(Q_{d_j}z + Q_{u_l}(1-z)\big) \frac{m_{d_j}}{m_{\tilde e_{L_{i}}}^2} \tilde{C}_0\Big( 1 , \frac{m_W^2}{m_{\tilde e_{L_{i}}}^2} , \frac{m_{d_j}^2}{(1-z) m_{\tilde e_{L_{i}}}^2} + \frac{m_{u_l}^2}{z m_{\tilde e_{L_{i}}}^2} \Big)  \Bigg] ~.
 \label{eq:lk f W}
\end{align}
In the equations presented above and in those that follow, $Q_f$ ($f = \ell, q$) denotes the electric charge in units of the (positive) elementary charge $e$, $n_c = 3$ represents the number of colors, and $\alpha_f$ ($f = \ell, q$) refers to the weak vector couplings of the fermions with the $Z$ boson. Note that the electric charges and weak couplings are universal across all generations:
\begin{equation}
Q_e = -1~,\quad Q_d = -\frac{1}{3}~, \quad Q_u = \frac{2}{3} ~, \quad \alpha_\ell = \frac{4 s_W^2 - 1}{4 s_W c_W} ~, \quad \alpha_d = \frac{4 s_W^2 - 3}{12 s_W c_W} ~,
\end{equation}
where $s_W = \sin\theta_W$ and $c_W = \cos\theta_W$ with the weak mixing angle $\theta_W$. 
However, we prefer to retain the flavor subscripts $Q_{f_i}$, $\alpha_{f_i}$ in the equations to indicate from which vertex in the diagram the corresponding couplings are introduced.

Moreover, in this work, to facilitate the identification of the placement of two RPV couplings appearing in the Barr--Zee diagrams, we note that $\mathrm{Im}(\lambda^{(\prime)}_a \lambda^{(\prime)*}_b) = -\mathrm{Im}(\lambda^{(\prime)}_b \lambda^{(\prime)*}_a)$, indicating that the two forms are not independent. We therefore adopt the convention that the coupling without complex conjugation, as used in the expressions of our results, is assigned to the vertex in the inner loop, while the complex-conjugated coupling corresponds to the vertex at the base of the Barr--Zee diagram.

The photon and $Z$ boson contributions are very well approximated by the leading logarithmic terms shown explicitly in Eqs.~\eqref{eq:lk f gamma} and \eqref{eq:lk f Z}. The terms with the Davydychev--Tausk function can be neglected to a very good approximation. In the $W$ contribution, Eq.~\eqref{eq:lk f W}, the Passarino-Veltman function $\tilde{C}_0$ appears, which is given by
\begin{equation}
    \tilde{C}_0(1,a,b) = \frac{a\log a}{(1-a)(a-b)} + \frac{b\log b}{(1-b)(b-a)} ~.
\end{equation}

A well-known feature of the Barr--Zee contribution with the $Z$ boson is its accidental suppression by the factor $1 - 4 s_W^2 \simeq 0.07$.
Among the contributions given above, the Barr--Zee contributions with the photon are generically the largest, as they are enhanced by very large logarithms. A resummation of such logs requires renormalization group techniques, which is beyond the scope of this paper. Inside those logarithms, the down quark mass and the strange quark mass should be replaced by appropriate hadronic scales, e.g., the pion and kaon masses.

Analogously to the lepton EDMs, we find for EDMs and chromo-EDMs of down-type quarks $d_k$: 
\begin{align}
 \frac{d_{d_k}^{f,\gamma}}{e} = \frac{\alpha_{\rm em}}{16 \pi^3} \sum_{i,j} \Bigg[ &\text{Im}\big(\lambda_{ijj} \lambda_{ikk}^{\prime \,*} \big) Q^2_{e_j}Q_{d_k}  \frac{m_{e_j}}{m_{\tilde \nu_{i}}^2} \bigg( \log\big( \frac{m_{\tilde \nu_{i}}^2}{m_{e_j}^2} \big) - 2 + \frac{m_{e_j}^2}{m_{\tilde \nu_{i}}^2} \Phi\big( \frac{m_{e_j}^2}{m_{\tilde \nu_{i}}^2} \big)\bigg) \notag\\ 
 + &\text{Im}\big(\lambda^\prime_{ijj} \lambda_{ikk}^{\prime \,*} \big) n_c Q^2_{d_j}Q_{d_k} \frac{m_{d_j}}{m_{\tilde \nu_{i}}^2} \bigg( \log\big( \frac{m_{\tilde \nu_{i}}^2}{m_{d_j}^2} \big) - 2 + \frac{m_{d_j}^2}{m_{\tilde \nu_{i}}^2} \Phi\big( \frac{m_{d_j}^2}{m_{\tilde \nu_{i}}^2} \big)\bigg)  \Bigg] ~, 
\label{eq:dk f gamma}
\end{align}
\begin{align}
 \frac{d_{d_k}^{f,Z}}{e} = &\frac{\alpha_{\rm em}}{16 \pi^3} \sum_{i,j}  \frac{1}{1 - m_Z^2/m_{\tilde \nu_{i}}^2}  \Bigg[ \notag\\
 &\text{Im}\big(\lambda_{ijj} \lambda_{ikk}^{\prime \,*} \big) Q_{e_j}\alpha_{e_j}\alpha_{d_k} \frac{m_{e_j}}{m_{\tilde \nu_{i}}^2} \bigg( \log\big( \frac{m_{\tilde \nu_{i}}^2}{m_Z^2} \big) - \frac{m_{e_j}^2}{m_Z^2} \Phi\big(\frac{m_{e_j}^2}{m_Z^2}\big) + \frac{m_{e_j}^2}{m_{\tilde \nu_{i}}^2} \Phi\big(\frac{m_{e_j}^2}{m_{\tilde \nu_{i}}^2}\big)\bigg) \notag\\
 + &\text{Im}\big(\lambda_{ijj}^\prime \lambda_{ikk}^{\prime \,*} \big) n_c Q_{d_j}\alpha_{d_j}\alpha_{d_k}  \frac{m_{d_j}}{m_{\tilde \nu_{i}}^2} \bigg( \log\big( \frac{m_{\tilde \nu_{i}}^2}{m_Z^2} \big) - \frac{m_{d_j}^2}{m_Z^2} \Phi\big(\frac{m_{d_j}^2}{m_Z^2}\big) + \frac{m_{d_j}^2}{m_{\tilde \nu_{i}}^2} \Phi\big(\frac{m_{d_j}^2}{m_{\tilde \nu_{i}}^2}\big)\bigg)   \Bigg]   ~,
\label{eq:dk f Z}
\end{align}
\begin{align}
 \frac{d_{d_k}^{f,W}}{e} =& -\frac{\alpha_{\rm em}}{128 \pi^3 s_W^2} \int_0^1 dz ~\Bigg[ \sum_{i,j} \text{Im}\big(\lambda_{ijj} \lambda_{ikk}^{\prime \, *} \big) Q_{e_j} z \frac{m_{e_j}}{m_{\tilde e_{L_{i}}}^2} \tilde{C}_0\Big( 1 , \frac{m_W^2}{m_{\tilde e_{L_{i}}}^2} , \frac{m_{e_j}^2}{(1-z) m_{\tilde e_{L_{i}}}^2} \Big) \notag\\ 
 + \sum_{i,j,l,n}  &\text{Im}\big(\lambda_{inj}^\prime \lambda_{ikk}^{\prime \, *} V_{lj} V_{ln}^* \big) n_c\big(Q_{d_j}z + Q_{u_l}(1-z)\big) \frac{m_{d_j}}{m_{\tilde e_{L_{i}}}^2} \tilde{C}_0\Big( 1 , \frac{m_W^2}{m_{\tilde e_{L_{i}}}^2} , \frac{m_{d_j}^2}{(1-z) m_{\tilde e_{L_{i}}}^2} + \frac{m_{u_l}^2}{z m_{\tilde e_{L_{i}}}^2} \Big)  \Bigg] ~,
\label{eq:dk f W}
\end{align}
\begin{equation}
 \frac{\tilde d_{d_k}^{f,g}}{g_s} = \frac{\alpha_s}{32 \pi^3} \sum_{i,j} \text{Im}\big(\lambda^\prime_{ijj} \lambda_{ikk}^{\prime \,*} \big)  \frac{m_{d_j}}{m_{\tilde \nu_{i}}^2} \bigg( \log\big( \frac{m_{\tilde \nu_{i}}^2}{m_{d_j}^2} \big) - 2 + \frac{m_{d_j}^2}{m_{\tilde \nu_{i}}^2} \Phi\big( \frac{m_{d_j}^2}{m_{\tilde \nu_{i}}^2} \big)\bigg) ~.
\label{eq:dk f g}
\end{equation}
The photon, $Z$ boson, and $W$ boson contributions are analogous in structure to the corresponding contributions to the charged lepton EDMs presented above.
The contribution to the chromo EDM~\eqref{eq:dk f g} comes from Barr--Zee diagrams with gluon exchange and a quark loop.
We note that in the $W$ contribution~\eqref{eq:dk f W} we have neglected the mass of the top quark that can appear in the main fermion line of the diagram. Neglecting the top mass is well justified as it would lead to a correction of the order of $m_t^2/m_{\tilde e_{L_i}}^2$, which is at most a few percent, given the current LHC constraints on slepton masses~\cite{Dreiner:2023bvs}. Our results agree with the ones given in Refs.~\cite{Yamanaka:2012hm, Yamanaka:2012ep} who used the same approximation.

\bigskip
%%%%%%%%%%%%%%%%%%%%%%%%%%%%%%%%%%%%%%%%%%%%%%%%
\paragraph{Barr--Zee diagrams with sfermions:} In addition to the Barr--Zee diagrams with fermions discussed above, there are the corresponding diagrams with sfermion loops. See examples in Fig.~\ref{fig:BarrZee_sfermion}. Summing the contributions from left-handed and right-handed sfermion loops, we obtain the following contributions to the EDMs of charged leptons:
\begin{align}
 \frac{d_{e_k}^{\tilde f,\gamma}}{e} =& - \frac{\alpha_{\rm em}}{32 \pi^3} \sum_{i,j} \Bigg[ \notag\\
 \text{Im}&\big(\lambda_{ijj} \lambda_{ikk}^* \big) Q^2_{e_j}Q_{e_k} \frac{m_{e_j}}{m_{\tilde \nu_{i}}^2} \bigg( \log\big( \frac{m_{\tilde \nu_{i}}^4}{m_{\tilde e_{L_j}}^2 m_{\tilde e_{R_j}}^2} \big) - 4 
 + \frac{m_{\tilde e_{L_j}}^2}{m_{\tilde \nu_{i}}^2} \Phi\big( \frac{m_{\tilde e_{L_j}}^2}{m_{\tilde \nu_{i}}^2} \big) + \frac{m_{\tilde e_{R_j}}^2}{m_{\tilde \nu_{i}}^2} \Phi\big( \frac{m_{\tilde e_{R_j}}^2}{m_{\tilde \nu_{i}}^2} \big) \bigg) \notag\\ 
 + \text{Im}&\big(\lambda_{ijj}^\prime \lambda_{ikk}^* \big) n_c Q^2_{d_j}Q_{e_k} \frac{m_{d_j}}{m_{\tilde \nu_{i}}^2} \bigg( \log\big( \frac{m_{\tilde \nu_{i}}^4}{m_{\tilde d_{L_j}}^2 m_{\tilde d_{R_j}}^2} \big) - 4 
 + \frac{m_{\tilde d_{L_j}}^2}{m_{\tilde \nu_{i}}^2} \Phi\big( \frac{m_{\tilde d_{L_j}}^2}{m_{\tilde \nu_{i}}^2} \big) + \frac{m_{\tilde d_{R_j}}^2}{m_{\tilde \nu_{i}}^2} \Phi\big( \frac{m_{\tilde d_{R_j}}^2}{m_{\tilde \nu_{i}}^2} \big) \bigg)  \Bigg] ~, 
\label{eq:lk sf gamma}
\end{align}
\begin{align}
 \frac{d_{e_k}^{\tilde f,Z}}{e} &= - \frac{\alpha_{\rm em}}{32 \pi^3} \sum_{i,j}  \frac{1}{1 - m_Z^2/m_{\tilde \nu_{i}}^2}  \Bigg[ \notag\\
 &\text{Im}\big(\lambda_{ijj} \lambda_{ikk}^* \big) Q_{e_j} \alpha_{e_k} \frac{m_{e_j}}{m_{\tilde \nu_{i}}^2} \bigg[\alpha_{\tilde{e}_{L_j}} \Big( \log\big( \frac{m_{\tilde \nu_{i}}^2}{m_Z^2} \big) - \frac{m_{\tilde e_{L_j}}^2}{m_Z^2} \Phi\big(\frac{m_{\tilde e_{L_j}}^2}{m_Z^2}\big) + \frac{m_{\tilde e_{L_j}}^2}{m_{\tilde \nu_{i}}^2} \Phi\big(\frac{m_{\tilde e_{L_j}}^2}{m_{\tilde \nu_{i}}^2}\big)\Big) \notag\\
 &\hspace{9em}+ \alpha_{\tilde{e}_{R_j}} \Big( \log\big( \frac{m_{\tilde \nu_{i}}^2}{m_Z^2} \big) - \frac{m_{\tilde e_{R_j}}^2}{m_Z^2} \Phi\big(\frac{m_{\tilde e_{R_j}}^2}{m_Z^2}\big) + \frac{m_{\tilde e_{R_j}}^2}{m_{\tilde \nu_{i}}^2} \Phi\big(\frac{m_{\tilde e_{R_j}}^2}{m_{\tilde \nu_{i}}^2}\big)\Big) \bigg] \notag\\
 + &\text{Im}\big(\lambda_{ijj}^\prime \lambda_{ikk}^* \big) n_c Q_{d_j} \alpha_{e_k} \frac{m_{d_j}}{m_{\tilde \nu_{i}}^2} \bigg[\alpha_{\tilde{d}_{L_j}} \Big( \log\big( \frac{m_{\tilde \nu_{i}}^2}{m_Z^2} \big) - \frac{m_{\tilde d_{L_j}}^2}{m_Z^2} \Phi\big(\frac{m_{\tilde d_{L_j}}^2}{m_Z^2}\big) + \frac{m_{\tilde d_{L_j}}^2}{m_{\tilde \nu_{i}}^2} \Phi\big(\frac{m_{\tilde d_{L_j}}^2}{m_{\tilde \nu_{i}}^2}\big)\Big) \notag\\
 &\hspace{9em}+ \alpha_{\tilde{d}_{R_j}} \Big( \log\big( \frac{m_{\tilde \nu_{i}}^2}{m_Z^2} \big) - \frac{m_{\tilde d_{R_j}}^2}{m_Z^2} \Phi\big(\frac{m_{\tilde d_{R_j}}^2}{m_Z^2}\big) + \frac{m_{\tilde d_{R_j}}^2}{m_{\tilde \nu_{i}}^2} \Phi\big(\frac{m_{\tilde d_{R_j}}^2}{m_{\tilde \nu_{i}}^2}\big)\Big) \bigg] \Bigg]   ~,
\label{eq:lk sf Z}
\end{align}
\begin{align}
 \frac{d_{e_k}^{\tilde f,W}}{e} =& \frac{\alpha_{\rm em}}{128 \pi^3 s_W^2} \int_0^1 dz ~\Bigg[ \sum_{i,j} \text{Im}\big(\lambda_{ijj} \lambda_{ikk}^* \big)  Q_{e_j}z  \frac{m_{e_j}}{m_{\tilde e_{L_{i}}}^2} \tilde{C}_0\Big( 1 , \frac{m_W^2}{m_{\tilde e_{L_{i}}}^2} , \frac{m_{\tilde e_{L_j}}^2}{(1-z) m_{\tilde e_{L_{i}}}^2} + \frac{m_{\tilde \nu_{j}}^2}{z m_{\tilde e_{L_{i}}}^2} \Big) \notag\\ 
 + \sum_{i,j,l,n}  &\text{Im}\big(\lambda_{inj}^\prime \lambda_{ikk}^* V_{lj} V_{ln}^* \big) n_c (Q_{d_j}z + Q_{u_l}(1-z)) \frac{m_{d_j}}{m_{\tilde e_{L_{i}}}^2} \tilde{C}_0\Big( 1 , \frac{m_W^2}{m_{\tilde e_{L_{i}}}^2} , \frac{m_{\tilde d_{L_j}}^2}{(1-z) m_{\tilde e_{L_{i}}}^2} + \frac{m_{\tilde u_{L_l}}^2}{z m_{\tilde e_{L_{i}}}^2} \Big)  \Bigg] ~.
\label{eq:lk sf W}
\end{align}
Our results agree with the ones presented in Refs.~\cite{Yamanaka:2012zq, Yamanaka:2012ep}. We note that here it is not a good approximation to neglect the terms with the Davydychev--Tausk function, as they are not suppressed by small mass ratios.
The above expressions for the $Z$ boson contribution contain the weak couplings of the left-handed and right-handed sleptons and down squarks, which are given by
\begin{equation}
 \alpha_{\tilde e_L} = \frac{2 s_W^2 - 1}{2 c_W s_W}~,\quad \alpha_{\tilde e_R} = \frac{s_W}{c_W}~,\quad  \alpha_{\tilde d_L} = \frac{2 s_W^2 - 3}{6 c_W s_W}~,\quad \alpha_{\tilde d_R} = \frac{s_W}{3 c_W} ~.
\end{equation}

For the corresponding expressions for the down quark EDMs and chromo-EDMs, we find, in agreement with Refs.~\cite{Yamanaka:2012zq, Yamanaka:2012ep},
\begin{align}
 \frac{d_{d_k}^{\tilde f,\gamma}}{e} =& - \frac{\alpha_{\rm em}}{32 \pi^3} \sum_{i,j} \Bigg[ \notag\\
 \text{Im}&\big(\lambda_{ijj} \lambda_{ikk}^{\prime \,*} \big) Q_{e_j}^2 Q_{d_k} \frac{m_{e_j}}{m_{\tilde \nu_{i}}^2} \bigg( \log\big( \frac{m_{\tilde \nu_{i}}^4}{m_{\tilde e_{L_j}}^2 m_{\tilde e_{R_j}}^2} \big) - 4 + \frac{m_{\tilde e_{L_j}}^2}{m_{\tilde \nu_{i}}^2} \Phi\big( \frac{m_{\tilde e_{L_j}}^2}{m_{\tilde \nu_{i}}^2} \big) + \frac{m_{\tilde e_{R_j}}^2}{m_{\tilde \nu_{i}}^2} \Phi\big( \frac{m_{\tilde e_{R_j}}^2}{m_{\tilde \nu_{i}}^2} \big) \bigg) \notag\\ 
 + \text{Im}&\big(\lambda_{ijj}^\prime \lambda_{ikk}^{\prime \,*} \big) n_c Q_{d_j}^2 Q_{d_k} \frac{m_{d_j}}{m_{\tilde \nu_{i}}^2} \bigg( \log\big( \frac{m_{\tilde \nu_{i}}^4}{m_{\tilde d_{L_j}}^2 m_{\tilde d_{R_j}}^2} \big) - 4 + \frac{m_{\tilde d_{L_j}}^2}{m_{\tilde \nu_{i}}^2} \Phi\big( \frac{m_{\tilde d_{L_j}}^2}{m_{\tilde \nu_{i}}^2} \big) + \frac{m_{\tilde d_{R_j}}^2}{m_{\tilde \nu_{i}}^2} \Phi\big( \frac{m_{\tilde d_{R_j}}^2}{m_{\tilde \nu_{i}}^2} \big) \bigg)  \Bigg] ~, 
\label{eq:dk sf gamma}
\end{align}
\begin{align}
 \frac{d_{d_k}^{\tilde f,Z}}{e} &= -\frac{\alpha_{\rm em}}{32 \pi^3} \sum_{i,j} \frac{1}{1 - m_Z^2/m_{\tilde \nu_{i}}^2} \Bigg[ \notag\\
 &\text{Im}\big(\lambda_{ijj} \lambda_{ikk}^{\prime \,*} \big) Q_{e_j} \alpha_{d_k} \frac{m_{e_j}}{m_{\tilde \nu_{i}}^2} \bigg[\alpha_{\tilde{e}_{L_j}} \Big( \log\big( \frac{m_{\tilde \nu_{i}}^2}{m_Z^2} \big) - \frac{m_{\tilde e_{L_j}}^2}{m_Z^2} \Phi\big(\frac{m_{\tilde e_{L_j}}^2}{m_Z^2}\big) + \frac{m_{\tilde e_{L_j}}^2}{m_{\tilde \nu_{i}}^2} \Phi\big(\frac{m_{\tilde e_{L_j}}^2}{m_{\tilde \nu_{i}}^2}\big)\Big) \notag\\
 &\hspace{9em}+ \alpha_{\tilde{e}_{R_j}} \Big( \log\big( \frac{m_{\tilde \nu_{i}}^2}{m_Z^2} \big) - \frac{m_{\tilde e_{R_j}}^2}{m_Z^2} \Phi\big(\frac{m_{\tilde e_{R_j}}^2}{m_Z^2}\big) + \frac{m_{\tilde e_{R_j}}^2}{m_{\tilde \nu_{i}}^2} \Phi\big(\frac{m_{\tilde e_{R_j}}^2}{m_{\tilde \nu_{i}}^2}\big)\Big) \bigg] \notag\\
 + &\text{Im}\big(\lambda_{ijj}^\prime \lambda_{ikk}^{\prime \,*} \big) n_c Q_{d_j} \alpha_{d_k} \frac{m_{d_j}}{m_{\tilde \nu_{i}}^2} \bigg[\alpha_{\tilde{d}_{L_j}} \Big( \log\big( \frac{m_{\tilde \nu_{i}}^2}{m_Z^2} \big) - \frac{m_{\tilde d_{L_j}}^2}{m_Z^2} \Phi\big(\frac{m_{\tilde d_{L_j}}^2}{m_Z^2}\big) + \frac{m_{\tilde d_{L_j}}^2}{m_{\tilde \nu_{i}}^2} \Phi\big(\frac{m_{\tilde d_{L_j}}^2}{m_{\tilde \nu_{i}}^2}\big)\Big) \notag\\
 &\hspace{9em}+ \alpha_{\tilde{d}_{R_j}} \Big( \log\big( \frac{m_{\tilde \nu_{i}}^2}{m_Z^2} \big) - \frac{m_{\tilde d_{R_j}}^2}{m_Z^2} \Phi\big(\frac{m_{\tilde d_{R_j}}^2}{m_Z^2}\big) + \frac{m_{\tilde d_{R_j}}^2}{m_{\tilde \nu_{i}}^2} \Phi\big(\frac{m_{\tilde d_{R_j}}^2}{m_{\tilde \nu_{i}}^2}\big)\Big) \bigg] \Bigg]   ~,
\label{eq:dk sf Z}
\end{align}
\begin{align}
 \frac{d_{d_k}^{\tilde f,W}}{e} =& \frac{\alpha_{\rm em}}{128 \pi^3 s_W^2} \int_0^1 dz~ \Bigg[ \sum_{i,j} \text{Im}\big(\lambda_{ijj} \lambda_{ikk}^{\prime \,*} \big) Q_{e_j} z \frac{m_{e_j}}{m_{\tilde e_{L_{i}}}^2} \tilde{C}_0\Big( 1 , \frac{m_W^2}{m_{\tilde e_{L_{i}}}^2} , \frac{m_{\tilde e_{L_j}}^2}{(1-z) m_{\tilde e_{L_{i}}}^2} + \frac{m_{\tilde \nu_{j}}^2}{z m_{\tilde e_{L_{i}}}^2} \Big) \notag\\ 
 + \sum_{i,j,l,n}  &\text{Im}\big(\lambda_{inj}^\prime \lambda_{ikk}^{\prime \,*} V_{lj} V_{ln}^* \big) n_c (Q_{d_j}z + Q_{u_l}(1-z)) \frac{m_{d_j}}{m_{\tilde e_{L_{i}}}^2} \tilde{C}_0\Big( 1 , \frac{m_W^2}{m_{\tilde e_{L_{i}}}^2} , \frac{m_{\tilde d_{L_j}}^2}{(1-z) m_{\tilde e_{L_{i}}}^2} + \frac{m_{\tilde u_{L_l}}^2}{z m_{\tilde e_{L_{i}}}^2} \Big)  \Bigg] ~,
\label{eq:dk sf W}
\end{align}
\begin{align}
 \frac{\tilde d_{d_k}^{\tilde f,g}}{g_s} = -\frac{\alpha_s}{64 \pi^3} \sum_{i,j}& \text{Im}\big(\lambda_{ijj}^\prime \lambda_{ikk}^{\prime \,*} \big) \notag\\
 \times &\frac{m_{d_j}}{m_{\tilde \nu_{i}}^2} \bigg( \log\big( \frac{m_{\tilde \nu_{i}}^4}{m_{\tilde d_{L_j}}^2 m_{\tilde d_{R_j}}^2} \big) - 4 + \frac{m_{\tilde d_{L_j}}^2}{m_{\tilde \nu_{i}}^2} \Phi\big( \frac{m_{\tilde d_{L_j}}^2}{m_{\tilde \nu_{i}}^2} \big) + \frac{m_{\tilde d_{R_j}}^2}{m_{\tilde \nu_{i}}^2} \Phi\big( \frac{m_{\tilde d_{R_j}}^2}{m_{\tilde \nu_{i}}^2} \big) \bigg)  \Bigg] ~.
\label{eq:dk sf g}
\end{align}
Analogous to the $W$ contribution with a fermion loop, we have neglected the mass of the top quark in Eq.~\eqref{eq:dk sf W}.
In Eqs.~\eqref{eq:lk sf gamma}-\eqref{eq:dk sf g}, although no fermions appear in the first loop, the fermion masses still enter the expressions through the trilinear $R$-parity violating interactions of the scalars [see Eq.~\eqref{eq:scalar int1} and Eq.~\eqref{eq:scalar int2}]. Also note that since the relative magnitudes of the slepton, squark, and sneutrino masses are unknown, the expansion of the Davydychev--Tausk $\Phi$ function is no longer applicable. Similarly, the order of terms in the expressions no longer necessarily reflects the hierarchy of their magnitudes.

\bigskip
%%%%%%%%%%%%%%%%%%%%%%%%%%%%%%%%%%%%%%%%%%%%%%%%
\paragraph{Barr--Zee diagrams with neutralinos, charginos, and gluinos:} Another class of diagrams are those that contain neutralinos, charginos, and gluinos instead of the SM gauge bosons. Example diagrams are depicted in Fig.~\ref{fig:gauginos}. These diagrams have been evaluated in Ref.~\cite{Yamanaka:2012qn}. The corresponding expressions for the EDMs and cEDMs are independent from the ones discussed so far, as they depend on additional parameters, in particular the gaugino masses and $\tan\beta$ -- the ratio of the two Higgs vacuum expectation values. It was found in Ref.~\cite{Yamanaka:2012qn} that the contributions involving the neutralinos, charginos, and gluinos might become relevant in the large $\tan\beta$ regime. We leave a detailed re-evaluation of such contributions for future work.

\bigskip
%%%%%%%%%%%%%%%%%%%%%%%%%%%%%%%%%%%%%%%%%%%%%%%%
\paragraph{Diagrams with 4 RPV couplings:} 

In addition to the diagrams considered in our analysis, which involve two RPV couplings, there also exist contributions from diagrams with four RPV couplings. Partial estimates for some of these diagrams were presented in Ref.~\cite{Abel:1999yz}, but to our knowledge, no comprehensive calculation has been carried out to date. While many RPV couplings are tightly constrained by existing data, some remain relatively unconstrained and could, in principle, be $O(1)$ in size. As a result, although contributions from four-coupling diagrams are generally expected to be subdominant, it is not excluded that they could yield non-negligible effects on EDMs in certain regions of parameter space. A complete study of these diagrams, accounting for all possible topologies and coupling combinations, represents a challenging but worthwhile task, which we leave for future work.

%%%%%%%%%%%%%%%%%%%%%%%%%%%%%%%%%%%%%%%%%%%%%%%%%%%
\section{Numerical Analysis} \label{sec:numerics}
%%%%%%%%%%%%%%%%%%%%%%%%%%%%%%%%%%%%%%%%%%%%%%%%%%%

After summarizing the results of our re-evaluation of the various Barr--Zee contributions to the EDMs and cEDMs in the previous section, we now confront these RPV contributions with the existing experimental bounds on EDMs. This gives us updated constraints on the RPV couplings.
Before presenting our results, we briefly summarize the numerical input that we use for the CKM matrix. The values of all the other physical parameters appearing in the formulas are taken from the PDG~\cite{ParticleDataGroup:2022pth}.

%%%%%%%%%%%%%%%%%%%%%%%%%%%%%%%%%%%%%%%%%%%%%%%%%%%
\subsection{CKM Input} \label{sec:CKM}
%%%%%%%%%%%%%%%%%%%%%%%%%%%%%%%%%%%%%%%%%%%%%%%%%%%

The $W$ Barr--Zee contributions to the EDMs involve CKM matrix elements, and we therefore specify our numerical approach.
We determine the CKM matrix elements from the input given by the PDG. In particular, we use~\cite{ParticleDataGroup:2022pth}
\begin{equation} \label{eq:CKM_input}
    |V_{cb}| = (40.8 \pm 1.4)\times 10^{-3}~,~~~ |V_{ub}| = (3.82 \pm 0.20)\times 10^{-3}~,~~~ \gamma = 65.9^\circ \pm 3.5^\circ~.
\end{equation}
The quoted values for $|V_{cb}|$ and $|V_{ub}|$ are averages of determinations using inclusive and exclusive tree-level $B$ decays.
For the sine of the Cabibbo angle, we use $\lambda \simeq 0.225$~\cite{ ParticleDataGroup:2022pth} and neglect its tiny uncertainty.

Working in the standard phase convention for the CKM matrix and expanding in powers of $\lambda$, all CKM entries can be expressed in terms of the above four input parameters using unitarity (see e.g. Ref.~\cite{Altmannshofer:2021uub}):
\begin{align} \label{eq:CKM}
V_{ud} &\simeq 1 - \frac{\lambda^2}{2} ~,& V_{us} &\simeq \lambda ~, &  V_{ub} &\simeq |V_{ub}| e^{-i \gamma} ~, \nonumber \\
V_{cd} &\simeq - \lambda ~,& V_{cs} &\simeq 1 - \frac{\lambda^2}{2} ~, & V_{cb} &= |V_{cb}| ~, \nonumber \\
V_{td} &\simeq |V_{cb}|\lambda - |V_{ub}|e^{i \gamma} \left(1-\frac{\lambda^2}{2}\right)~,& V_{ts} &\simeq -|V_{cb}|\left(1-\frac{\lambda^2}{2}\right) - |V_{ub}|\lambda e^{i \gamma} ~, &  V_{tb} &\simeq 1 ~.
\end{align}
For each CKM element, the expressions take into account relative $\mathcal O(\lambda^2)$ corrections, and they are therefore accurate at the few per mille level. Numerically, this procedure gives
%
%\begin{equation}
% \begin{pmatrix} 
% V_{ud} & V_{us} & V_{ub} \\
% V_{cd} & V_{cs} & V_{cb} \\
% V_{td} & V_{ts} & V_{tb}
% \end{pmatrix} \simeq \begin{pmatrix} 
% 0.975 & 0.225 & (1.56 - i 3.49) \cdot 10^{-3} \\
% -0.225 & 0.975 & 4.08 \cdot 10^{-2} \\
% (7.66 - i 3.40) \cdot 10^{-3} & - (4.01 + i 0.08) \cdot 10^{-2} & 1
% \end{pmatrix} ~.
%\end{equation}
%
\begin{equation}
 V_\text{CKM} \simeq \begin{pmatrix} 
 0.975 & 0.225 & (1.56 - i 3.49) \times 10^{-3} \\
 -0.225 & 0.975 & 4.08 \times 10^{-2} \\
 (7.66 - i 3.40) \times 10^{-3} & - (4.01 + i 0.08) \times 10^{-2} & 1
 \end{pmatrix} ~.
\end{equation}

As the EDMs are proportional to the imaginary parts of combinations of RPV couplings and CKM matrix elements, it is important to keep the imaginary parts in the CKM elements.

%%%%%%%%%%%%%%%%%%%%%%%%%%%%%%%%%%%%%%%%%%%%%%%%%%%
\subsection{Constraints on the RPV couplings} \label{sec:Constraints}
%%%%%%%%%%%%%%%%%%%%%%%%%%%%%%%%%%%%%%%%%%%%%%%%%%%

Using the previously provided formulas for the lepton and quark EDMs, Eqs.~\eqref{eq:lk f gamma}-\eqref{eq:dk sf g}, we present the constraints on specific combinations of RPV couplings from the experimental data of the electron and neutron EDMs collected in Section~\ref{sec:exp}. Table~\ref{tab:eEDMconstraints} and Table~\ref{tab:nEDMconstraints} detail the upper bounds on various RPV coupling combinations from the electron EDM and the neutron EDM, respectively. The second columns of the tables indicate the corresponding non-zero Barr--Zee type contributions, arising from different gauge boson exchanges.

To obtain the shown constraints, we set the masses of all SUSY particles to $1$\,TeV and switch on one coupling combination at a time. The constraints, therefore, hold assuming the absence of accidental cancellations among several contributions. Although the experimental results for muon and tau EDMs can, in principle, constrain additional combinations of RPV couplings, we find that given the current experimental sensitivities (see Table~\ref{tab:EDM_exp}), the resulting bounds are not meaningful. Specifically, the strongest upper limits derived from the muon EDM appear for combinations of two RPV couplings at the $\mathcal{O}(10^4)$ level, while those from the tau EDM reach the $\mathcal{O}(10^6)$ level.

\renewcommand{\arraystretch}{1.3}
%%%%%%%%%%%%%%%%%%%%%%%%%%%%%%%%%%%%
\begin{table}[tbh]
\begin{ruledtabular}
\begin{tabular}{ccc}
RPV couplings                                         & Contributions            & Limits from $d_e$\\ \hline
$|\mathrm{Im}(\lambda_{233} \lambda_{211}^*)|$        & $d_e^\gamma+d_e^Z+d_e^W$ & $7.98 \times 10^{-7}$\\
$|\mathrm{Im}(\lambda_{322} \lambda_{311}^*)|$        & $d_e^\gamma+d_e^Z+d_e^W$ & $9.09 \times 10^{-6}$                              \\
$|\mathrm{Im}(\lambda^\prime_{i11} \lambda_{i11}^*)|\ (i=2,3)$ & $d_e^\gamma+d_e^Z+d_e^W$ & $5.72 \times 10^{-4}$                             \\
$|\mathrm{Im}(\lambda^\prime_{i12} \lambda_{i11}^*)|\ (\mathrm{with\ Re}(\lambda^\prime_{i12} \lambda_{i11}^*)=0,\ i=2,3)$ &          $d_e^W$         & $0.117$                             \\
$|\mathrm{Re}(\lambda^\prime_{i12} \lambda_{i11}^*)|\ (\mathrm{with\ Im}(\lambda^\prime_{i12} \lambda_{i11}^*)=0,\ i=2,3)$ &          $d_e^W$         & $3.59$                 \\
$|\mathrm{Im}(\lambda^\prime_{i13} \lambda_{i11}^*)|\ (\mathrm{with\ Re}(\lambda^\prime_{i13} \lambda_{i11}^*)=0,\ i=2,3)$ &          $d_e^W$         & $5.27 \times 10^{-4}$                        \\
$|\mathrm{Re}(\lambda^\prime_{i13} \lambda_{i11}^*)|\ (\mathrm{with\ Im}(\lambda^\prime_{i13} \lambda_{i11}^*)=0,\ i=2,3)$ &          $d_e^W$         & $1.18 \times 10^{-3}$                       \\
$|\mathrm{Im}(\lambda^\prime_{i21} \lambda_{i11}^*)|\ (\mathrm{with\ Re}(\lambda^\prime_{i21} \lambda_{i11}^*)=0,\ i=2,3)$ &          $d_e^W$         & $2.33$         \\
$|\mathrm{Re}(\lambda^\prime_{i21} \lambda_{i11}^*)|\ (\mathrm{with\ Im}(\lambda^\prime_{i21} \lambda_{i11}^*)=0,\ i=2,3)$ &          $d_e^W$         & $71.7$            \\
$|\mathrm{Im}(\lambda^\prime_{i22} \lambda_{i11}^*)|\ (i=2,3)$ & $d_e^\gamma+d_e^Z+d_e^W$ & $4.21 \times 10^{-5}$      \\
$|\mathrm{Im}(\lambda^\prime_{i23} \lambda_{i11}^*)|\ (\mathrm{with\ Re}(\lambda^\prime_{i23} \lambda_{i11}^*)=0,\ i=2,3)$ &          $d_e^W$         & $1.00 \times 10^{-4}$          \\
$|\mathrm{Re}(\lambda^\prime_{i23} \lambda_{i11}^*)|\ (\mathrm{with\ Im}(\lambda^\prime_{i23} \lambda_{i11}^*)=0,\ i=2,3)$ &          $d_e^W$         & $5.13 \times 10^{-3}$         \\
$|\mathrm{Im}(\lambda^\prime_{i31} \lambda_{i11}^*)|\ (\mathrm{with\ Re}(\lambda^\prime_{i31} \lambda_{i11}^*)=0,\ i=2,3)$ &          $d_e^W$         & $0.476$       \\
$|\mathrm{Re}(\lambda^\prime_{i31} \lambda_{i11}^*)|\ (\mathrm{with\ Im}(\lambda^\prime_{i31} \lambda_{i11}^*)=0,\ i=2,3)$ &          $d_e^W$         & $1.06$         \\
$|\mathrm{Im}(\lambda^\prime_{i32} \lambda_{i11}^*)|\ (\mathrm{with\ Re}(\lambda^\prime_{i32} \lambda_{i11}^*)=0,\ i=2,3)$ &          $d_e^W$         & $4.53 \times 10^{-3}$      \\
$|\mathrm{Re}(\lambda^\prime_{i32} \lambda_{i11}^*)|\ (\mathrm{with\ Im}(\lambda^\prime_{i32} \lambda_{i11}^*)=0,\ i=2,3)$ &          $d_e^W$         & $0.231$        \\
$|\mathrm{Im}(\lambda^\prime_{i33} \lambda_{i11}^*)|\ (i=2,3)$ & $d_e^\gamma+d_e^Z+d_e^W$ & $1.50 \times 10^{-6}$ \\
\end{tabular}
\end{ruledtabular}
\caption{Constraints on the RPV couplings, with upper limits derived from the current experimental EDM measurements of the electron with all sfermion masses involved in the calculation specified to be 1 TeV. We also highlight, in the second column, the types of gauge bosons that appear in the Barr--Zee type diagrams, contributing to the electron EDM.}
\label{tab:eEDMconstraints}
\end{table}
%%%%%%%%%%%%%%%%%%%%%%%%%%%%%%%%%%%%
\begin{table}[tbh]
\begin{ruledtabular}
\begin{tabular}{ccc}
RPV couplings                                                    & Contributions            & Limits from $d_n$ \\ \hline
$|\mathrm{Im}(\lambda_{i22} \lambda_{i11}^{\prime *})|\ (i=1,3)$ & $d_n^\gamma+d_n^Z+d_n^W$ &        $0.146$ \\
$|\mathrm{Im}(\lambda_{i33} \lambda_{i11}^{\prime *})|\ (i=1,2)$ & $d_n^\gamma+d_n^Z+d_n^W$ &        $0.0120$ \\
$|\mathrm{Im}(\lambda_{i11} \lambda_{i11}^{\prime *})|\ (i=2,3)$ & $d_n^\gamma+d_n^Z+d_n^W$ &        $19.7$  \\
$|\mathrm{Im}(\lambda^\prime_{i12} \lambda^{\prime *}_{i11})|\ (\mathrm{with\ Re}(\lambda^\prime_{i12} \lambda_{i11}^{\prime *})=0,\ i=1,2,3)$ &          $d_n^W$         & $\mathcal{O}(100)$                             \\
$|\mathrm{Re}(\lambda^\prime_{i12} \lambda^{\prime *}_{i11})|\ (\mathrm{with\ Im}(\lambda^\prime_{i12} \lambda_{i11}^{\prime *})=0,\ i=1,2,3)$ &          $d_n^W$         & $\mathcal{O}(10^4)$                             \\
$|\mathrm{Im}(\lambda^\prime_{i13} \lambda^{\prime *}_{i11})|\ (\mathrm{with\ Re}(\lambda^\prime_{i13} \lambda_{i11}^{\prime *})=0,\ i=1,2,3)$ &          $d_n^W$         & $3.31$                             \\
$|\mathrm{Re}(\lambda^\prime_{i13} \lambda^{\prime *}_{i11})|\ (\mathrm{with\ Im}(\lambda^\prime_{i13} \lambda_{i11}^{\prime *})=0,\ i=1,2,3)$ &          $d_n^W$         & $7.39$                             \\
$|\mathrm{Im}(\lambda^\prime_{i21} \lambda^{\prime *}_{i11})|\ (\mathrm{with\ Re}(\lambda^\prime_{i21} \lambda_{i11}^{\prime *})=0,\ i=1,2,3)$ &          $d_n^W$         & $\mathcal{O}(10^4)$                             \\
$|\mathrm{Re}(\lambda^\prime_{i21} \lambda^{\prime *}_{i11})|\ (\mathrm{with\ Im}(\lambda^\prime_{i21} \lambda_{i11}^{\prime *})=0,\ i=1,2,3)$ &          $d_n^W$         & $\mathcal{O}(10^5)$                             \\
$|\mathrm{Im}(\lambda_{i22}^{\prime} \lambda_{i11}^{\prime *})|\ (i=1,2,3)$ & $d_n^\gamma+d_n^Z+d_n^W$ &        $0.875$          \\
$|\mathrm{Im}(\lambda^\prime_{i23} \lambda^{\prime *}_{i11})|\ (\mathrm{with\ Re}(\lambda^\prime_{i23} \lambda_{i11}^{\prime *})=0,\ i=1,2,3)$ &          $d_n^W$         & $0.630$                             \\
$|\mathrm{Re}(\lambda^\prime_{i23} \lambda^{\prime *}_{i11})|\ (\mathrm{with\ Im}(\lambda^\prime_{i23} \lambda_{i11}^{\prime *})=0,\ i=1,2,3)$ &          $d_n^W$         & $32.3$                             \\
$|\mathrm{Im}(\lambda^\prime_{i31} \lambda^{\prime *}_{i11})|\ (\mathrm{with\ Re}(\lambda^\prime_{i31} \lambda_{i11}^{\prime *})=0,\ i=1,2,3)$ &          $d_n^W$         & $\mathcal{O}(10^3)$                             \\
$|\mathrm{Re}(\lambda^\prime_{i31} \lambda^{\prime *}_{i11})|\ (\mathrm{with\ Im}(\lambda^\prime_{i31} \lambda_{i11}^{\prime *})=0,\ i=1,2,3)$ &          $d_n^W$         & $\mathcal{O}(10^3)$                             \\
$|\mathrm{Im}(\lambda^\prime_{i32} \lambda^{\prime *}_{i11})|\ (\mathrm{with\ Re}(\lambda^\prime_{i32} \lambda_{i11}^{\prime *})=0,\ i=1,2,3)$ &          $d_n^W$         & $28.5$                             \\
$|\mathrm{Re}(\lambda^\prime_{i32} \lambda^{\prime *}_{i11})|\ (\mathrm{with\ Im}(\lambda^\prime_{i32} \lambda_{i11}^{\prime *})=0,\ i=1,2,3)$ &          $d_n^W$         & $\mathcal{O}(10^3)$                             \\
$|\mathrm{Im}(\lambda_{i33}^{\prime} \lambda_{i11}^{\prime *})|\ (i=1,2,3)$ & $d_n^\gamma+d_n^Z+d_n^W$ &       $0.0179$ \\
\end{tabular}
\end{ruledtabular}
\caption{Constraints on the RPV couplings, with upper limits derived from the current experimental EDM measurements of the neutron with all sfermion masses involved in the calculation specified to be 1 TeV. We also highlight, in the second column, the types of gauge bosons that appear in the Barr--Zee type diagrams, contributing to the neutron EDM. Note that in some cases, the constraints are weaker than generic bounds from perturbativity considerations.}
\label{tab:nEDMconstraints}
\end{table}
%%%%%%%%%%%%%%%%%%%%%%%%%%%%%%%%%%%%
\renewcommand{\arraystretch}{1.}

Note that our analysis is conducted within the super CKM basis. While our expressions are consistent with the analytical results in Refs.~\cite{Yamanaka:2012hm,Yamanaka:2012zq,Yamanaka:2012ep}, there are some notable differences in the numerical constraints obtained for certain couplings, even when taking into account the impact of the latest experimental data in updating our constraints. These differences arise mainly from the different definitions of $\lambda^\prime$, or more specifically, from the CKM rotation that is present in Eq.~\eqref{eq:LQD} and Eq.~\eqref{eq:scalar int2}.

\begin{figure}[t!]
\centering
\hspace{2pt}
{\includegraphics[width=0.297\textwidth]{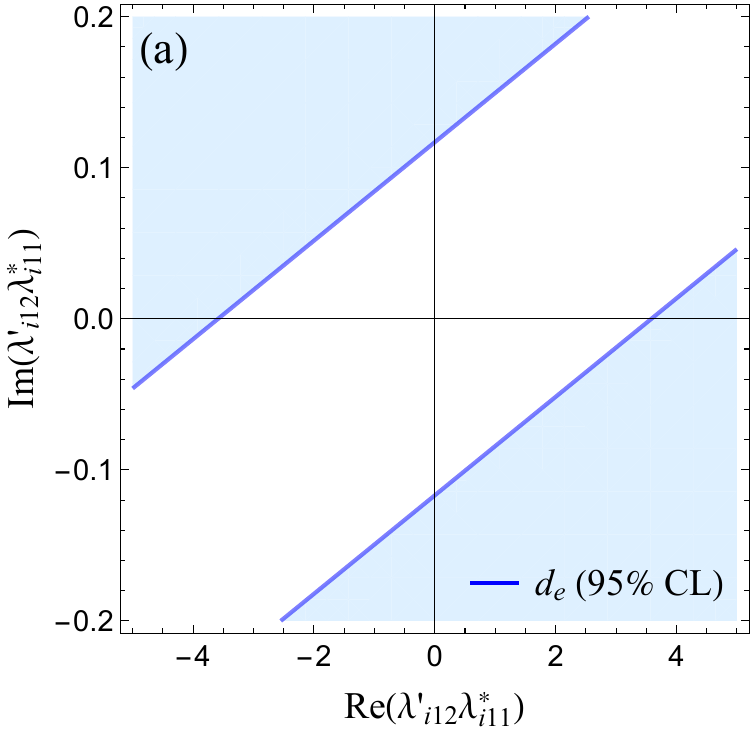}}
\hspace{0pt}
{\includegraphics[width=0.32\textwidth]{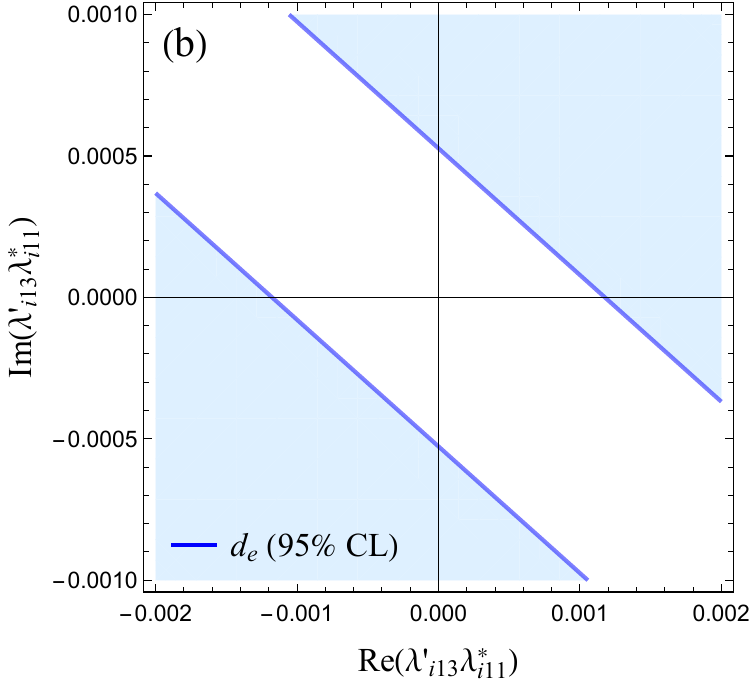}}
\hspace{3pt}
{\includegraphics[width=0.29\textwidth]{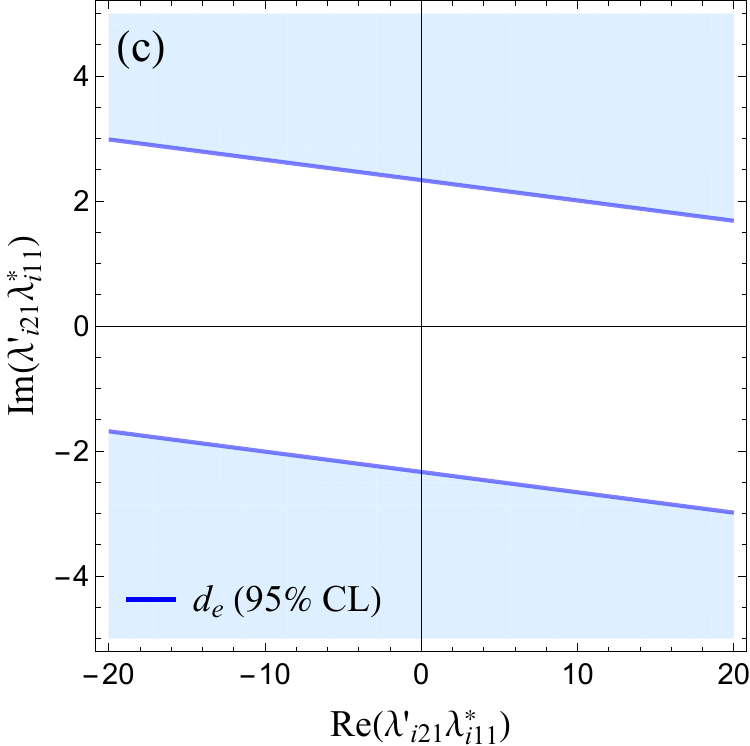}}
\\[15pt]
{\includegraphics[width=0.317\textwidth]{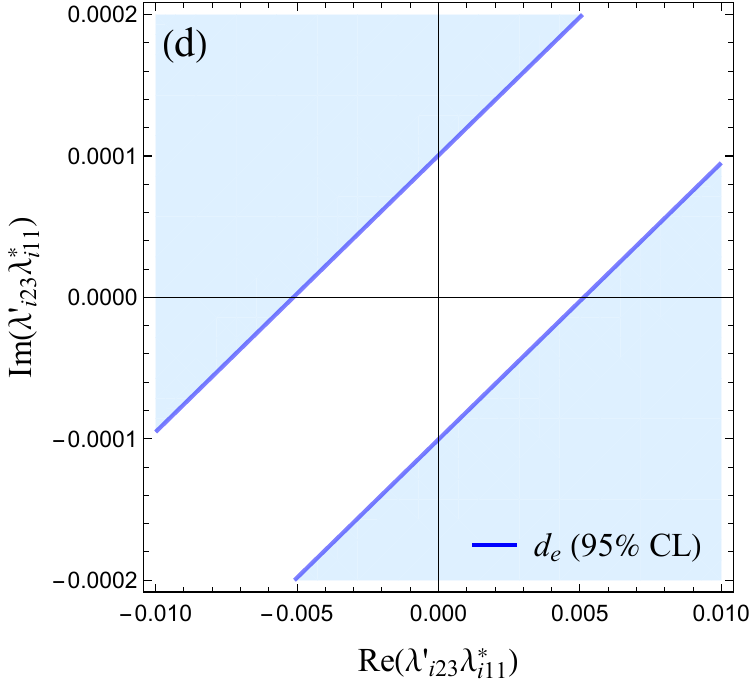}}
\hspace{2pt}
{\includegraphics[width=0.297\textwidth]{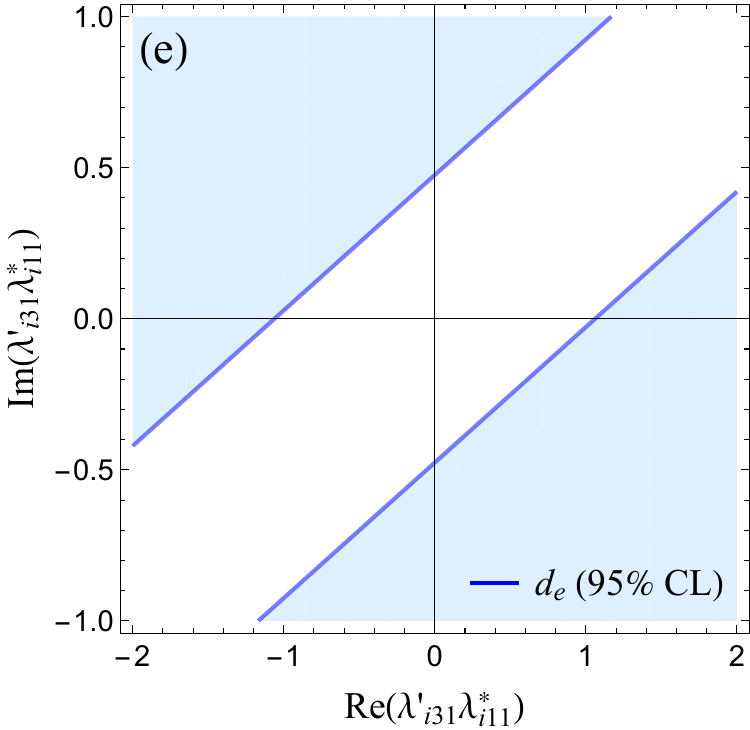}}
\hspace{0pt}
{\includegraphics[width=0.307\textwidth]{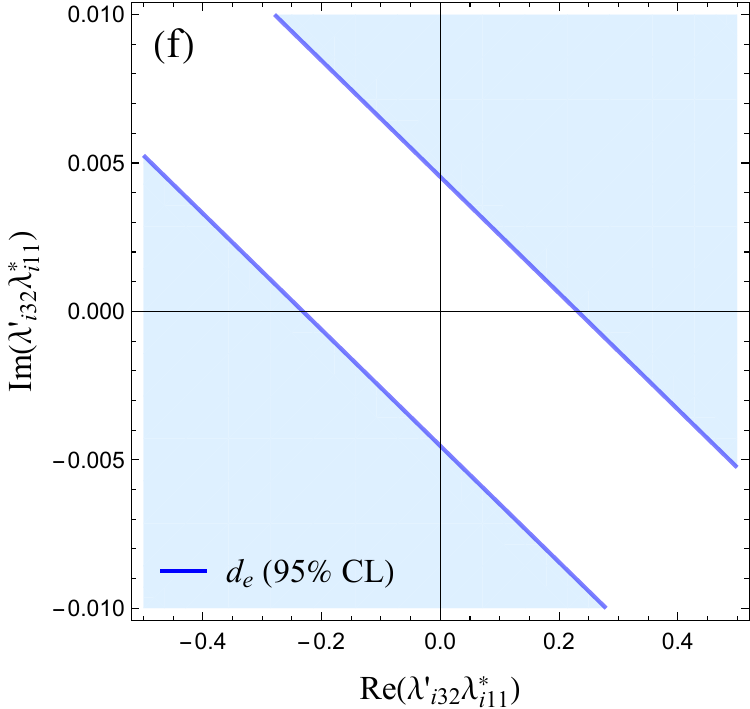}}
\\[15pt]
{\includegraphics[width=0.29\textwidth]{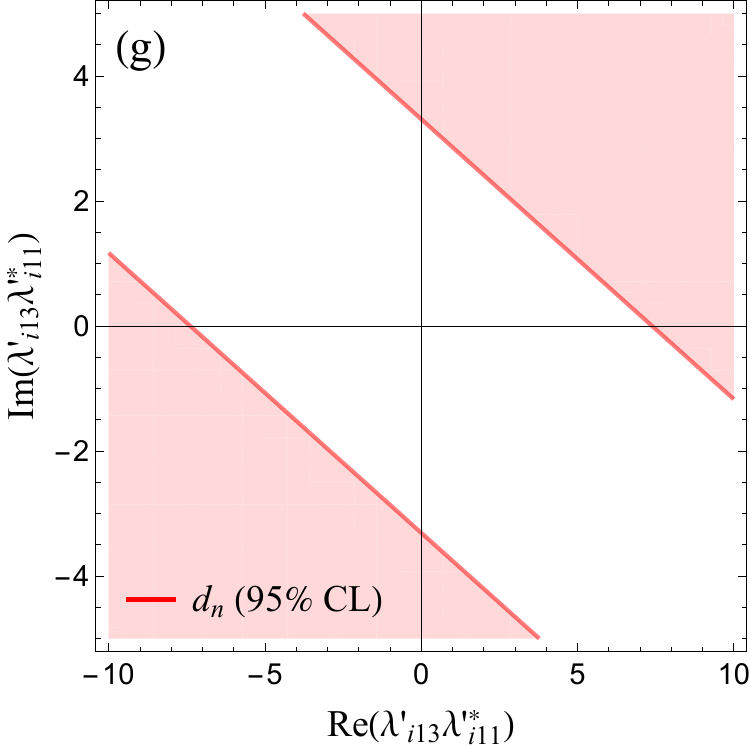}}
\hspace{10pt}
{\includegraphics[width=0.3\textwidth]{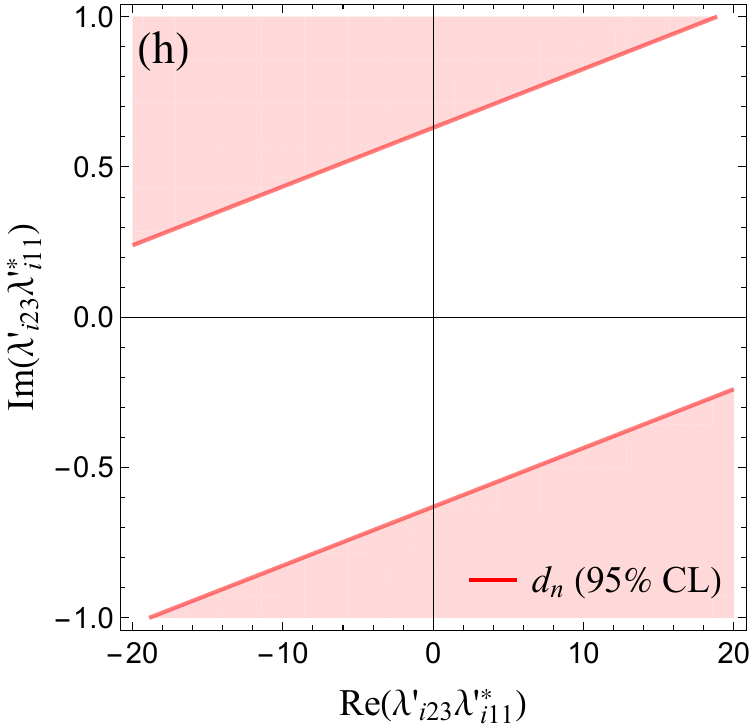}}
\caption{Constraints on combinations of RPV couplings in their complex plane, derived from the current 95\% CL experimental bounds on the electron (blue contours) and neutron (red contours) EDMs with all sfermion masses involved in the calculation specified to be 1 TeV. The shaded blue and red regions indicate parameter space disfavored by the respective EDM limits.}
\label{fig:contours}
\end{figure}

Additionally, we take into account the complex nature of the CKM matrix elements in our analysis. As a result, some coupling combinations cannot be expressed as simple upper bounds derived from the experimental EDM results. Instead, these cases correspond to an upper bound on a function of the real and imaginary parts of the product of two couplings. For simplicity, in Tables~\ref{tab:eEDMconstraints} and~\ref{tab:nEDMconstraints}, for the cases discussed above, we present the upper limits for the imaginary part when the real part is set to zero, and for the real part when the imaginary part is set to zero. Fig.~\ref{fig:contours} shows the complete constraints on these coupling combinations in the complex plane. The corresponding limits shown in Tables~\ref{tab:eEDMconstraints} and~\ref{tab:nEDMconstraints} correspond to the intersections of the respective contours with the real and imaginary axes. As clearly illustrated in Fig.~\ref{fig:contours}, the limits on the real (imaginary) parts shown in Tables~\ref{tab:eEDMconstraints} and~\ref{tab:nEDMconstraints} can become either stronger or weaker when the corresponding imaginary (real) part is nonzero.

Furthermore, in contrast to the results in Ref.~\cite{Yamanaka:2012ep}, it turns out that when working in the super CKM basis, the RPV couplings appearing at the base of the $W$-mediated Barr--Zee type diagrams must have equal second and third subindices even in the case of the $\lambda^\prime$-type couplings ($\lambda^\prime_{ikk}$). This is analogous in result to the case of $\lambda$-type couplings ($\lambda_{ikk}$), for which the equality of the second and third subindices reflects the assumption that flavor mixing in the lepton sector is negligible. Moreover, since the experimental constraints considered in this work originate from the EDMs of the electron or neutron, the subindex $k$ must be 1 in our results. Under the convention in this paper that the complex-conjugated coupling displayed in the results $\mathrm{Im}(\lambda^{(\prime)}_a \lambda^{(\prime)*}_b)$ corresponds to the vertex at the base of the Barr--Zee type diagram, the RPV couplings with complex conjugation appearing in Tables~\ref{tab:eEDMconstraints} and~\ref{tab:nEDMconstraints} must therefore take the form $\lambda^{(\prime)*}_{i11}$.

Complementary constraints on RPV couplings can, for example, be obtained from LHC searches, Higgs data, $B$ decays, meson oscillations, and other flavor observables~\cite{Domingo:2018qfg, Dreiner:2023bvs, Choudhury:2024ggy, Choudhury:2024yxd, Dreiner:2025kfd}. Different processes are sensitive to distinct combinations of RPV couplings and assumptions; therefore, a direct comparison is not straightforward.

%%%%%%%%%%%%%%%%%%%%%%%%%%%%%%%%%%%%
\begin{figure}[t!]
\centering
{\includegraphics[width=0.47\textwidth]{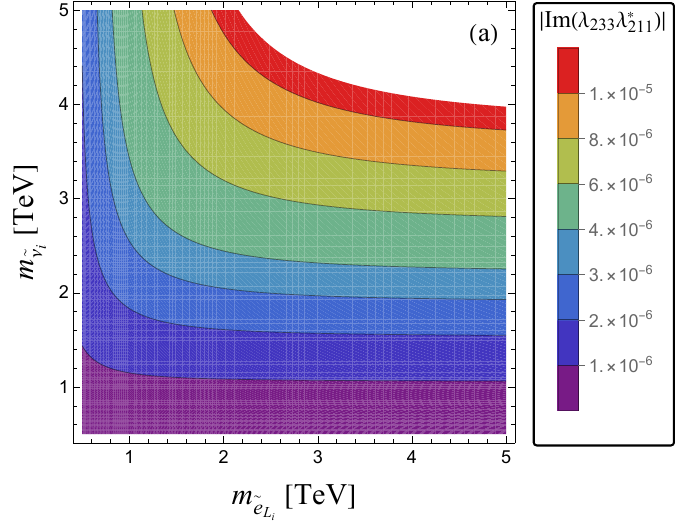}}
\hspace{10pt}
{\includegraphics[width=0.47\textwidth]{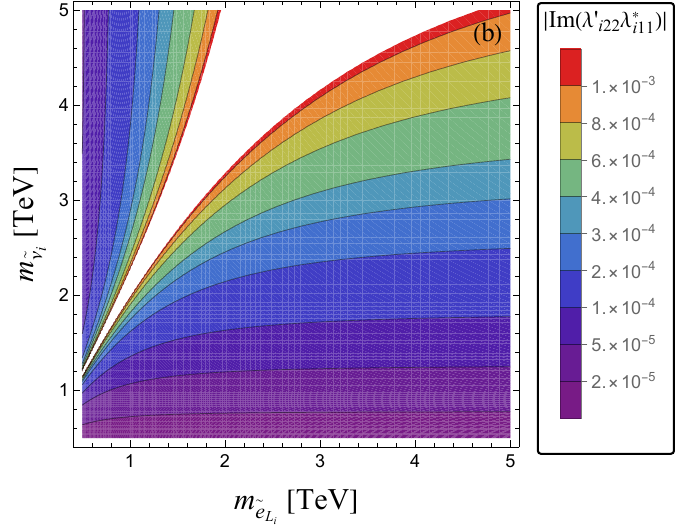}}
\\[15pt]
{\includegraphics[width=0.47\textwidth]{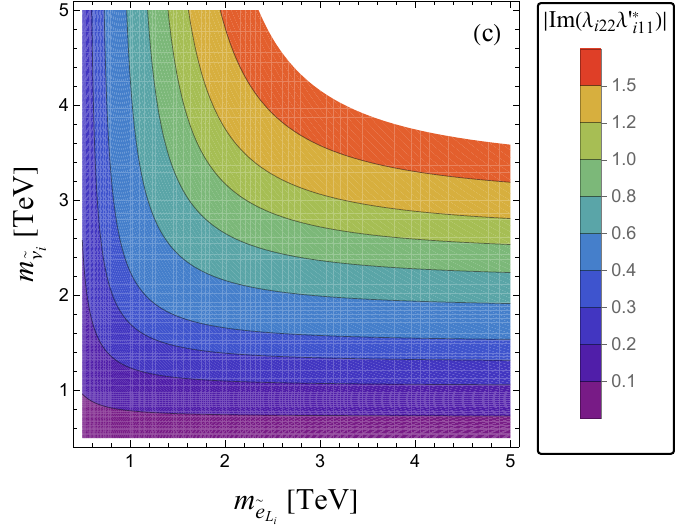}}
\hspace{13pt}
{\includegraphics[width=0.47\textwidth]{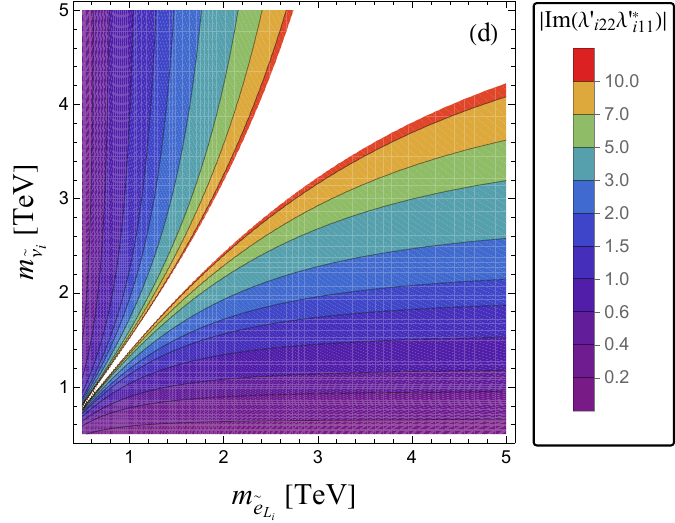}}
\caption{Contour plots illustrating the 95\% CL constraints on four representative combinations of RPV couplings as functions of the charged slepton and sneutrino masses. Panels (a) and (b) correspond to constraints derived from the electron EDM, while panels (c) and (d) show those from the neutron EDM. In all plots, the masses of other relevant supersymmetric particles in the first loop are assumed to be $1$~TeV.}
\label{fig:RPV_contours}
\end{figure}
%%%%%%%%%%%%%%%%%%%%%%%%%%%%%%%%%%%%

Since all possible constraints on combinations of RPV couplings from the electron and neutron EDMs can be computed analytically using the equations given in Sec.~\ref{sec:EDMs}, with the results depending on the masses of various supersymmetric particles entering the loop diagrams in a nontrivial way, we do not attempt to visualize the entire parameter space. Instead, Fig.~\ref{fig:RPV_contours} presents four representative coupling combinations\footnote{These combinations are chosen because, in each case, the EDM receives contributions from both the $\gamma/Z$- and $W$-exchange Barr--Zee type diagrams.} to illustrate the structure of the constraints as functions of the sneutrino and slepton masses, which are associated with the $\gamma/Z$- and $W$-mediated Barr--Zee type diagrams, respectively, due to their matching electric charges. Specifically, in the $\gamma/Z$-mediated diagrams, the sneutrino serves as the scalar propagator in the second loop, while in the $W$-mediated case, the corresponding scalar is the charged slepton.

In Fig.~\ref{fig:RPV_contours}, panel (a) shows the constraint on a pair of $\lambda$-type couplings from the electron EDM; panel (b) corresponds to a combination of $\lambda$ and $\lambda^\prime$ couplings, also from the electron EDM; panel (c) presents a constraint from the neutron EDM involving both $\lambda$ and $\lambda^\prime$ couplings; and panel (d) involves a pair of $\lambda^\prime$-type couplings constrained by the neutron EDM. The contours, corresponding to the 95\% CL upper limits, are displayed in the $(m_{\tilde{\ell}},m_{\tilde{\nu}})$ plane, where the vertical axis corresponds to the sneutrino mass and the horizontal axis to the charged slepton mass. While $\tilde{\nu}$ and $\tilde{\ell}$ are generally expected to be closely spaced in mass due to $SU(2)_L$ symmetry in the MSSM, we explore a broader range for illustration, varying both masses independently from 0.5 to 5~TeV. The masses of all other supersymmetric particles entering the first loop are fixed at 1~TeV.

We note that Ref.~\cite{Yamanaka:2012ep} pointed out that the $\gamma/Z$ and $W$ contributions enter with opposite signs, with the $\gamma$ contribution typically dominating under the assumption of a common SUSY mass of 1~TeV. We find that the $\gamma/Z$ contributions indeed dominate in the lower-right regions in panels (b) and (d) of Fig.~\ref{fig:RPV_contours}. In contrast, the upper-left regions of these panels are dominated by the $W$ contribution. The white intermediate region in panels (b) and (d) represents the parameter space where the two contributions approximately cancel, leading to a sharp weakening of the constraint. This cancellation, however, does not occur in panels (a) and (c) of Fig.~\ref{fig:RPV_contours}, where the $\gamma/Z$ and $W$ contributions add constructively due to their aligned signs.

In general, the relative importance of the $\gamma/Z$ and $W$ contributions depends sensitively on the mass hierarchy between the sneutrino and the charged slepton. A lighter sneutrino enhances the $\gamma/Z$ contribution, corresponding to the lower-right regions in all panels (a)-(d) of Fig.~\ref{fig:RPV_contours}, whereas a lighter slepton enhances the $W$ contribution, which dominates in the upper-left regions of these panels. Moreover, the sign correlation between the two contributions is determined by the RPV coupling entering the effective vertex in the inner loop of the Barr--Zee type diagrams. According to the convention adopted in this work, where the non-conjugated coupling in the notation ``$\mathrm{Im}(\lambda^{(\prime)} \lambda^{(\prime)*})$" corresponds to the one appearing in the inner loop, if this coupling is of $\lambda$ type [as in panels (a) and (c)], the $\gamma/Z$ and $W$ contributions have the same sign; whereas if it is of $\lambda^\prime$ type [as in panels (b) and (d)], they carry opposite signs.

By comparing Tables~\ref{tab:eEDMconstraints} and~\ref{tab:nEDMconstraints}, it is observed that the constraints derived from the electron EDM are generally stronger than those from the neutron EDM. However, for the combination of $\lambda_{i11}$ and $\lambda^\prime_{i11}$ couplings, the neutron EDM, despite providing weaker bounds in general, can lead to stronger constraints within certain regions of the parameter space. This is because, as for the case of $\lambda$--$\lambda^\prime$ combinations, the electron EDM constraint exhibits a cancellation region similar to Fig.~\ref{fig:RPV_contours}(b), which significantly weakens the sensitivity. In contrast, no such cancellation occurs in the neutron EDM case like Fig.~\ref{fig:RPV_contours}(c), enabling it to provide comparatively stronger bounds within this region of parameter space.

%%%%%%%%%%%%%%%%%%%%%%%%%%%%%%%%%%%%%%%%%%%%%%%%%%%
\subsection{Implications for the proton EDM} \label{sec:proton}
%%%%%%%%%%%%%%%%%%%%%%%%%%%%%%%%%%%%%%%%%%%%%%%%%%%

As already mentioned in Section~\ref{sec:EDMs}, the MSSM with the RPV interactions that we consider mainly generates EDMs for leptons and down-type quarks. The contributions to the up quark EDM and chromo-EDM are much smaller. 

This has interesting implications for the proton EDM. In scenarios in which the down quark EDM and chromo-EDM provide the dominant contribution to EDMs of nucleons, isospin symmetry gives the approximate relation [c.f. Eqs.~\eqref{eq:dn} and~\eqref{eq:dp}] 
\begin{equation} \label{eq:relation}
 d_p \simeq - \frac{1}{4} d_n ~. 
\end{equation}
Note that Eq.~\eqref{eq:relation} assumes that the proton and neutron EDMs are to a good approximation determined by the valence contributions of the up and down quarks as indicated by lattice calculations \cite{Bhattacharya:2016zcn, Gupta:2018lvp, Park:2025rxi}. However, if there is a substantial strange quark contribution~\cite{Vecchi:2025jbb}, the relation in Eq.~\eqref{eq:relation} would be modified. Similarly, sizable contributions from the Weinberg operator or from four-quark operators would also likely lead to modifications of the relation. See the discussion in Section~\ref{sec:exp}.

Taking Eq.~\eqref{eq:relation} at face value, the current experimental bound on the neutron EDM thus gives in our setup the indirect limit $|d_p| \lesssim 0.5 \times 10^{-26} e\, {\rm cm}$. 
If future experiments were to observe a proton EDM above this bound, or if they were to observe non-zero EDMs for both the neutron and proton but in violation of the relation in Eq.~\eqref{eq:relation}, the RPV contributions to EDMs discussed in this paper would not be sufficient for an explanation, and other BSM physics would be required.  

%%%%%%%%%%%%%%%%%%%%%%%%%%%%%%%%%%%%%%%%%%%%%%%%%%%
\section{Conclusions} \label{sec:conclusions}
%%%%%%%%%%%%%%%%%%%%%%%%%%%%%%%%%%%%%%%%%%%%%%%%%%%

In this paper, we have updated the constraints on the Minimal Supersymmetric Standard Model with trilinear $R$-parity violation using existing experimental limits on electric dipole moments. Our focus was on the $LQD$ and $LLE$ couplings. We do not consider $UDD$ couplings that, in combination with the $LQD$ and $LLE$ couplings, would be subject to extremely strong constraints from proton decay. 

As is well known, RPV contributions to EDMs of quarks and leptons, as well as chromo-EDMs of quarks, arise first at the two-loop level in the MSSM with RPV.
We have revisited the two-loop calculations of several classes of Barr--Zee type diagrams, which in many cases are expected to give the dominant contributions.
In particular, we have calculated the full set of Barr--Zee diagrams involving all SM gauge bosons (photons, $Z$ bosons, $W$ bosons, and gluons) with loops of fermions and sfermions.

We have performed our calculation in a generic $R_\xi$ background field gauge and have carefully monitored the effect of $\gamma_5$, making use of the \textquotesingle t~Hooft--Veltman scheme. Our final analytic results agree with existing calculations in the literature~\cite{Yamanaka:2012hm, Yamanaka:2012zq, Yamanaka:2012ep, Yamanaka:2012qn} and provide an important independent cross-check. We then confronted the two-loop EDM contributions with the current experimental limits on the electron and neutron EDMs to derive constraints on the RPV couplings.
Working in the super-CKM basis, we obtained constraints both on the imaginary parts of RPV coupling combinations and on some of the real parts. The constraints are summarized in Tables~\ref{tab:eEDMconstraints} and~\ref{tab:nEDMconstraints}. Some illustrative constraints as functions of the slepton masses are shown in Fig.~\ref{fig:RPV_contours}.

We point out that in the RPV framework considered here, there is a sharp correlation between the neutron and proton EDMs. As the contributions to the up quark EDM are much smaller than those to the down quark EDM, one expects that the proton EDM is a factor of four smaller than the neutron EDM and has the opposite sign. This testable prediction provides continued motivation for improving the searches of the neutron EDM and for performing direct searches of the proton EDM. 

As a final comment, we note that one may worry about the constraints on muon $g-2$ and muon EDM obtained from atoms and molecules. To address this concern, Ref.~\cite{DGS_25} analyzes data on $e^ + e^- \to \mu^+ \mu^-$ obtained from LEP, BaBar, Belle and Belle~II and also that experimental data plus the notion of ``optimized observables''~\cite{Atwood:1991ka}.

%%%%%%%%%%%%%%%%%%%%%%%%%%%%%%%%%%%%%%%%%%%%%%%%%%%
\section*{Acknowledgments} 
%%%%%%%%%%%%%%%%%%%%%%%%%%%%%%%%%%%%%%%%%%%%%%%%%%%
 
F.X. thanks Ding Yu Shao for pointing out the potential role of the axial anomaly. The research of W.A. is supported by the U.S. Department of Energy grant number DE-SC0010107. The work of B.D. was partly supported by the U.S. Department of Energy under grant No.~DE-SC0017987, and by a Humboldt Fellowship from the Alexander von Humboldt Foundation. The work of F.X. was partly supported by the U.S. Department of Energy under grant No.~DE-SC0017987, and by the National Science Foundations of China under Grant No.~12275052 and No.~12147101.

\begin{appendix}
%%%%%%%%%%%%%%%%%%%%%%%%%%%%%%%%%%%%%%%%%%%%%%%%%%%
\section{Details of the Two-Loop Calculation} \label{app:2loop}
%%%%%%%%%%%%%%%%%%%%%%%%%%%%%%%%%%%%%%%%%%%%%%%%%%%

While the final results of our analytical calculations are consistent with Refs.~\cite{Yamanaka:2012hm,Yamanaka:2012zq,Yamanaka:2012ep}, there are subtleties in the intermediate steps that deserve a careful discussion.

%%%%%%%%%%%%%%%%%%%%%%%%%%%%%%%
\subsection{Flavor basis and CKM matrix}
%%%%%%%%%%%%%%%%%%%%%%%%%%%%%%%

As mentioned in Section~\ref{sec:RPV}, the flavor basis adopted in our work is defined such that all charged fermions are rotated to their mass eigenstates, with the corresponding sfermions rotated in the same manner. In contrast, previous studies did not implement this rotation. Consequently, the interpretation of our $\lambda^\prime$ couplings differ from that in Refs.~\cite{Yamanaka:2012hm,Yamanaka:2012zq,Yamanaka:2012ep} by a CKM rotation, given by $\lambda^\prime_{ijk} \to \lambda^\prime_{imk} V^*_{jm}$. As discussed in Section~\ref{sec:numerics}, we also take into account the complex nature of the CKM matrix elements. In fact, it is these two differences that lead to the substantial variation between our numerical results in Tables~\ref{tab:eEDMconstraints} and ~\ref{tab:nEDMconstraints} and those of Refs.~\cite{Yamanaka:2012hm,Yamanaka:2012zq,Yamanaka:2012ep} in the constraints for $\lambda^\prime_{ijk}$ expressed in the same notation. Moreover, our approach allows us to derive additional constraints on the real parts of the RPV couplings (see Tables~\ref{tab:eEDMconstraints} and ~\ref{tab:nEDMconstraints}), with the source of CPV provided by the phase in the CKM matrix. To cross-check the reliability and consistency of our results, we used the old electron and neutron EDM experimental data~\cite{Hudson:2011zz,Baker:2006ts}, rotated the $\lambda^\prime$ in our expressions back to the definition used in Refs.~\cite{Yamanaka:2012hm,Yamanaka:2012zq,Yamanaka:2012ep}, and successfully reproduced all numerical results therein.

%%%%%%%%%%%%%%%%%%%%%%%%%%%%%%%
\subsection{Axial anomaly and treatment of $\gamma^5$}
%%%%%%%%%%%%%%%%%%%%%%%%%%%%%%%

In the calculation of the Barr--Zee type diagrams shown in Fig.~\ref{fig:BarrZee_fermion}, it is worth noting that the fermion loops are coupled to supersymmetric particles. These loops contain interaction vertices that explicitly involve chiral projection operators, see Eqs.~\eqref{eq:LLE} and \eqref{eq:LQD}, and consequently, $\gamma^5$. It therefore becomes necessary to consider the potential presence of axial anomalies when performing the calculation in dimensional regularization. 

As the final results for the EDMs are finite, it is common lore that calculations using a naively anti-commuting $\gamma^5$ should give the correct answer. However, to the best of our knowledge, previous studies, such as Refs.~\cite{Yamanaka:2012hm,Yamanaka:2012ep}, have not checked this explicitly. We carried out such cross checks adopting the \textquotesingle t~Hooft--Veltman scheme~\cite{tHooft:1972tcz} for $\gamma^5$. 

The structure of the fermion loops in Fig.~\ref{fig:BarrZee_fermion} exhibits a strong similarity to that of the well-known Adler--Bell--Jackiw (ABJ) triangle diagram~\cite{Adler:1969gk, Bell:1969ts}.
Specifically, the loop momentum in the diagram has components in all $4 - 2\epsilon$ dimensions, while $\gamma^5$ is defined in four dimensions. Unlike the usual anti-commutation relation, $\gamma^5$ commutes with the gamma matrices associated with the extra dimensions. As a result, when evaluating the trace, potential infinitesimal contributions may arise from the extra-dimensional components of the loop momentum and the associated gamma matrices. When combined with the divergences in the loop momentum integral, these infinitesimal terms could lead to additional finite contributions.

Nevertheless, we observe that such an anomalous contribution does not emerge in any of the cases we have considered in this paper, despite the similarity between the fermion loops in Fig.~\ref{fig:BarrZee_fermion} and the ABJ triangle diagram. This is attributed to the fact that, during the calculation, the number of gamma matrices appearing in the trace over the fermion loop is at least one less than that in the ABJ case. As a consequence, certain potentially infinitesimal terms vanish identically. From another perspective, the EDM should generally not be directly affected by axial anomalies, as it is not induced by chiral nonconservation in the same manner as processes like $\pi^0 \to \gamma \gamma$. This can also be seen from the fact that the anomalous contribution vanishes within each individual diagram, rather than being canceled through a summation over multiple diagrams.

%%%%%%%%%%%%%%%%%%%%%%%%%%%%%%%
\subsection{Barr--Zee Type diagrams with $W$ boson exchange}
%%%%%%%%%%%%%%%%%%%%%%%%%%%%%%%

%%%%%%%%%%%%%%%%%%%%%%%%%%%%%%%
\begin{figure}[tb]
\centering
\subfigure[]{
\includegraphics[width=0.32\textwidth]{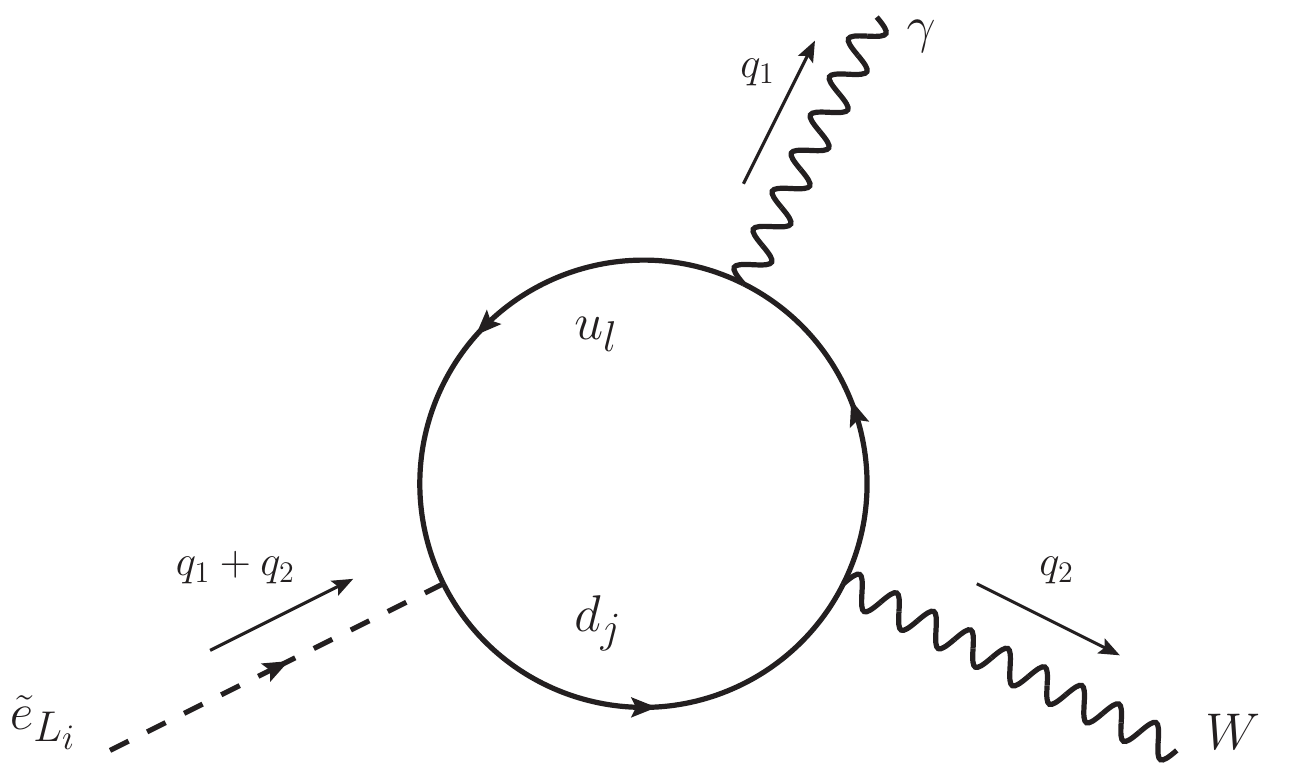}}
\subfigure[]{ \includegraphics[width=0.32\textwidth]{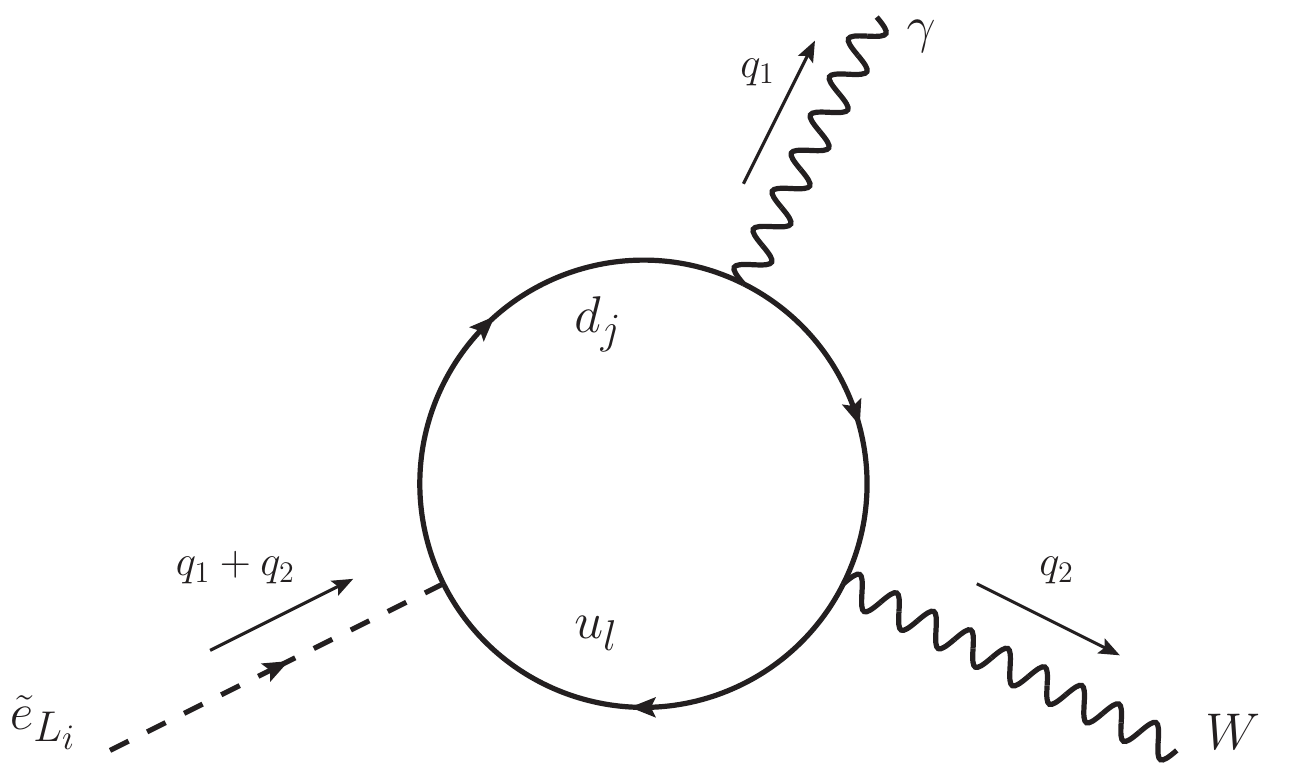}}
\subfigure[]{ \includegraphics[width=0.32\textwidth]{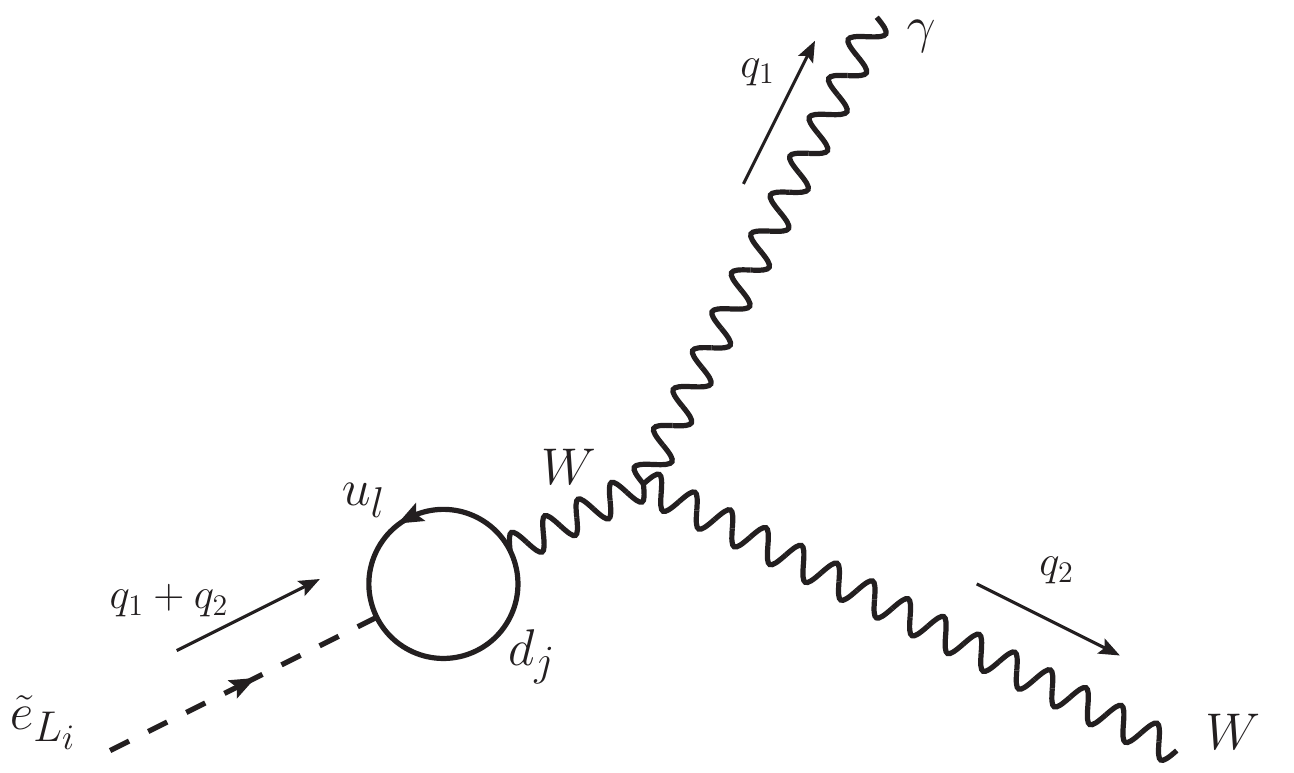}}
\subfigure[]{ \includegraphics[width=0.33\textwidth]{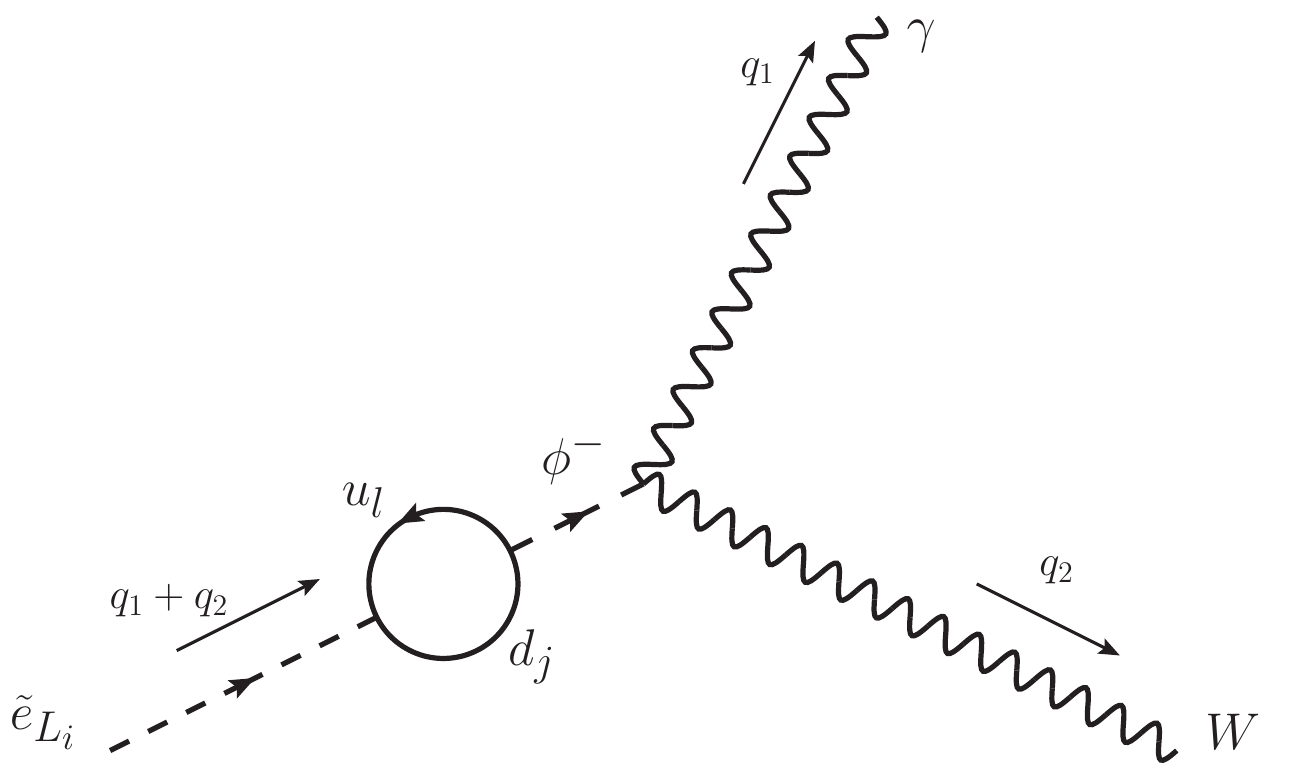}}
\subfigure[]{ \includegraphics[width=0.33\textwidth]{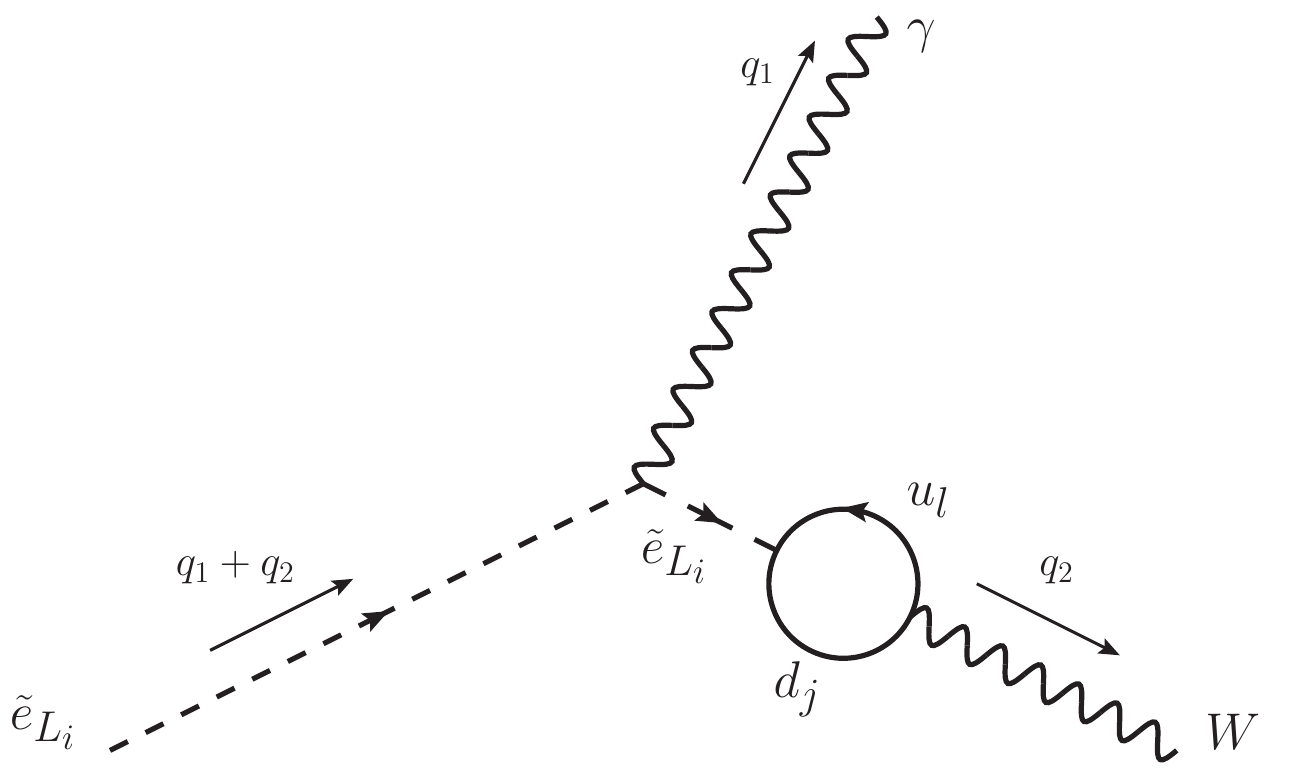}}
\caption{Effective $\tilde{e}_{L} \gamma W$ vertex with quark loops. The corresponding diagrams for lepton loops can be obtained by replacing the up-type quark with a neutrino and the down-type quark with a charged lepton in the figures. (Note that diagram (a) is absent in that case.) The $\phi^-$ in diagram (d) is the Nambu-Goldstone boson.}
\label{fig:1st_loop_fermion}
\end{figure}
%%%%%%%%%%%%%%%%%%%%%%%%%%%%%%%
\begin{figure}[tb]
\centering
\subfigure[]{
\includegraphics[width=0.32\textwidth]{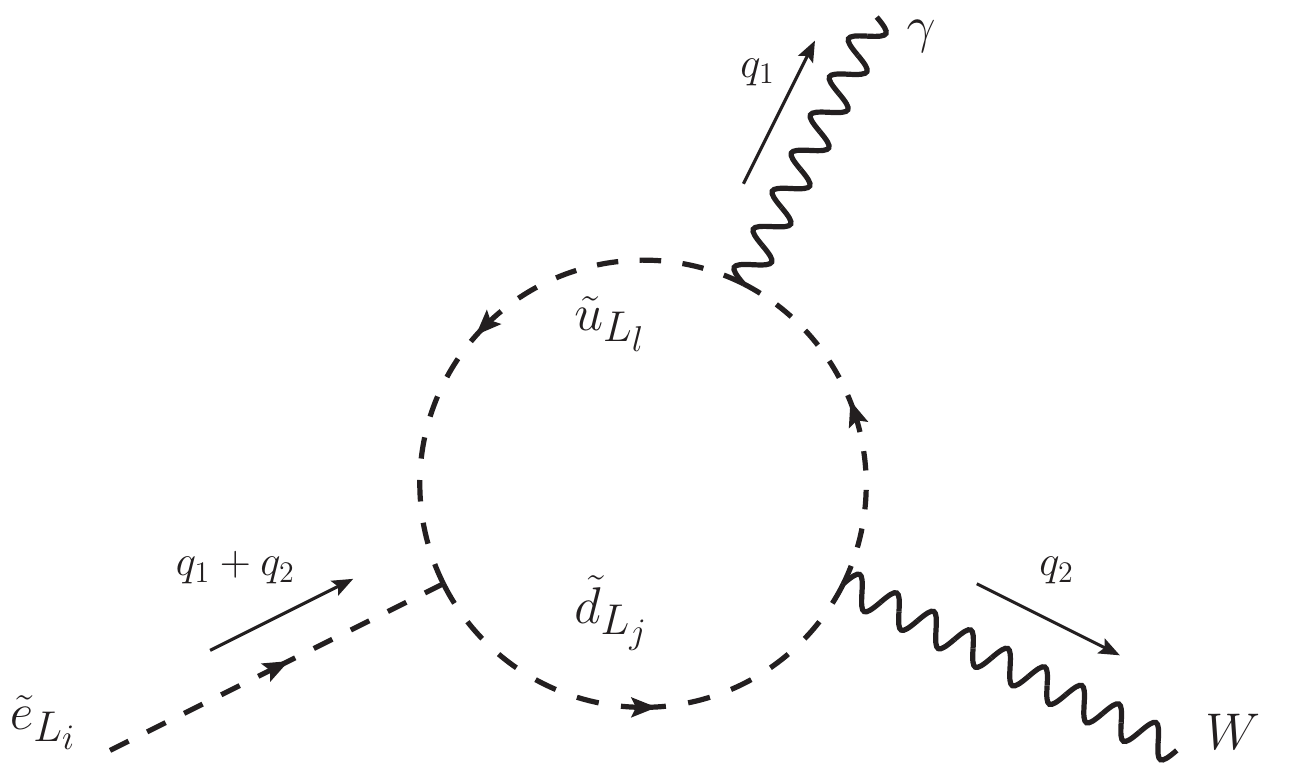}}
\subfigure[]{ \includegraphics[width=0.32\textwidth]{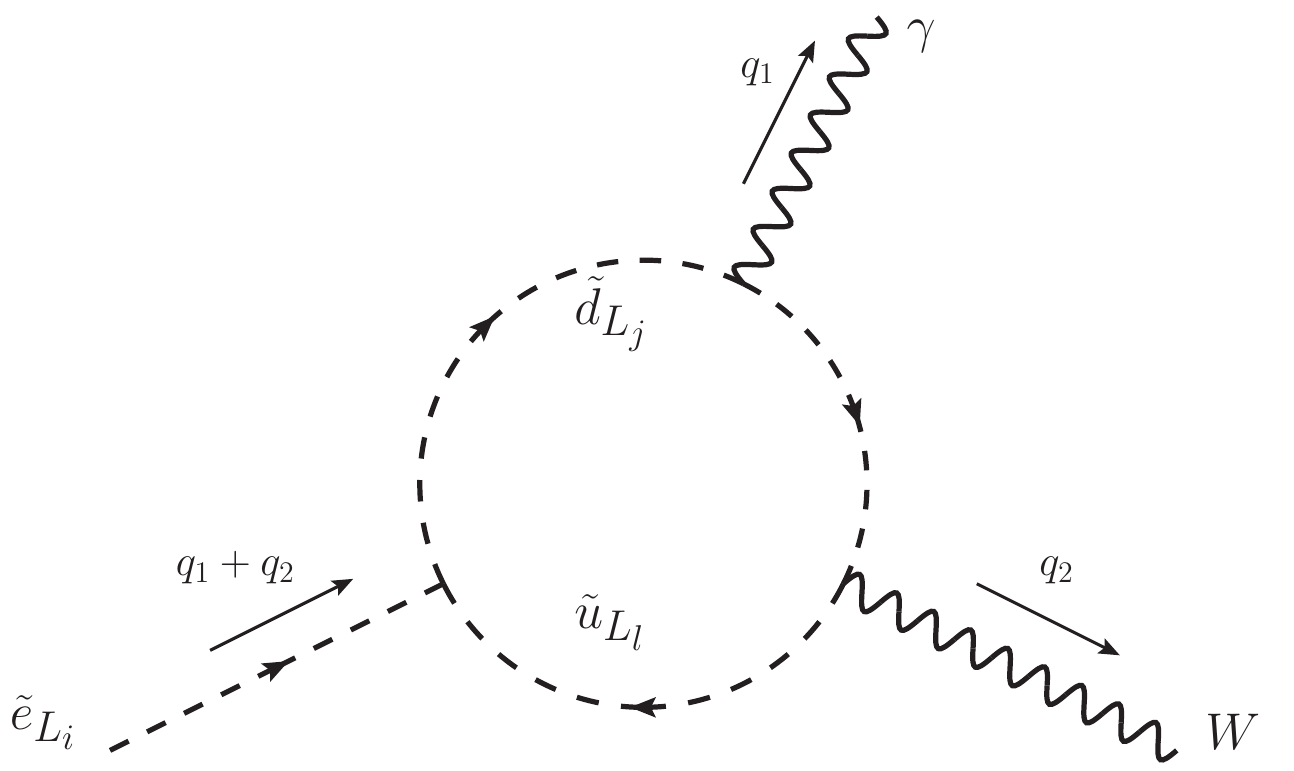}}
\subfigure[]{ \includegraphics[width=0.32\textwidth]{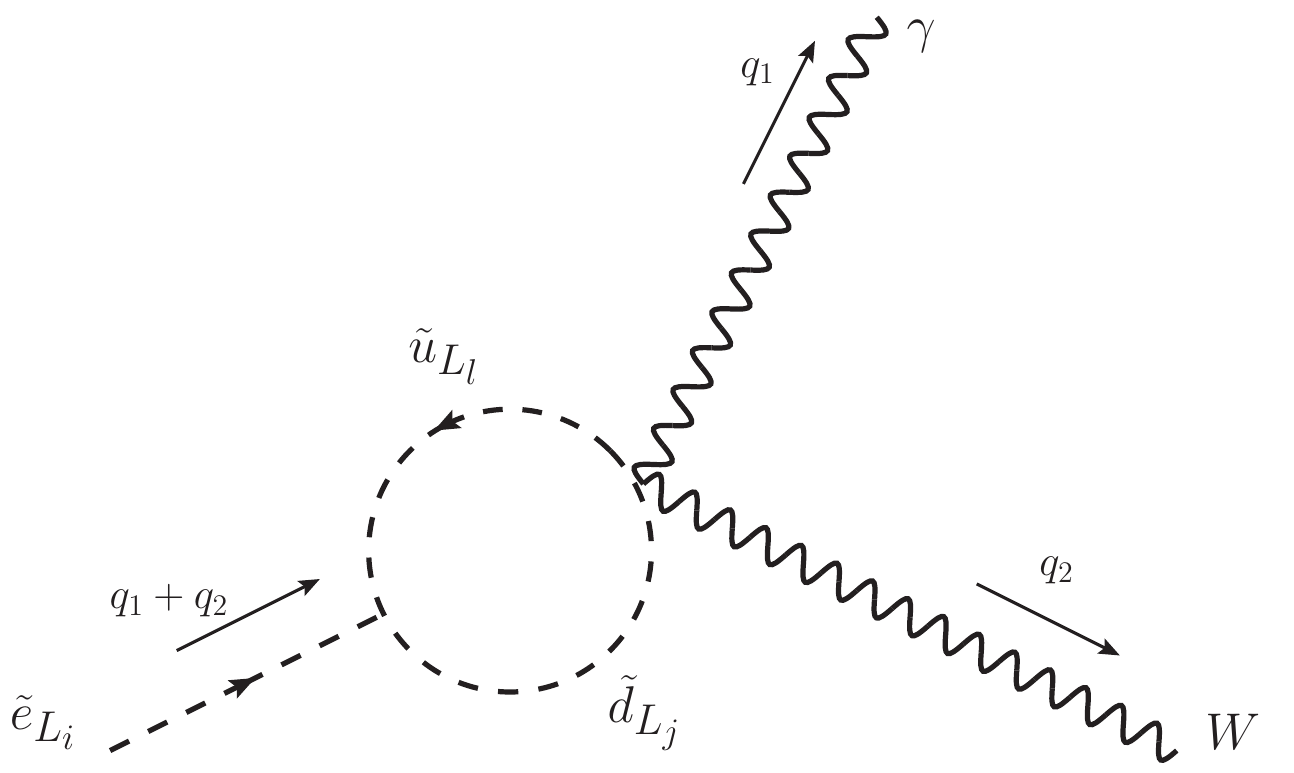}}
\subfigure[]{ \includegraphics[width=0.33\textwidth]{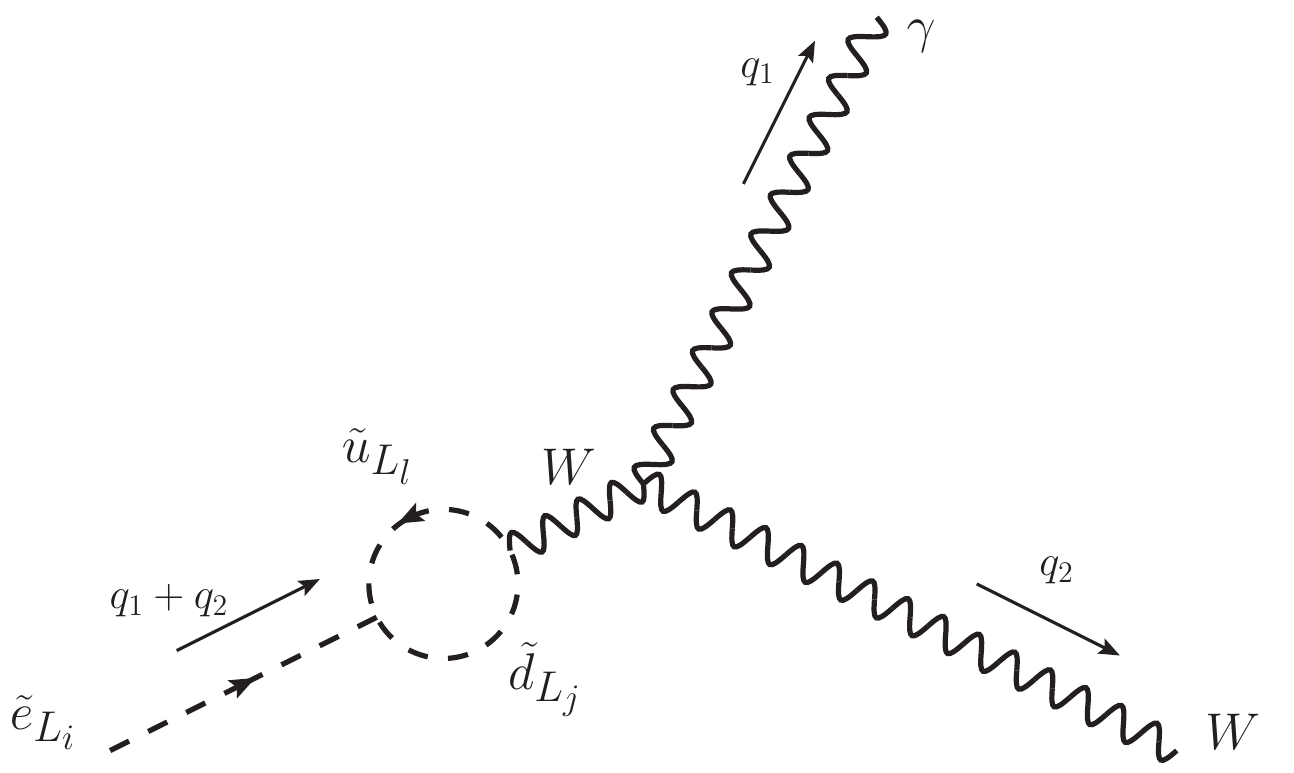}}
\subfigure[]{ \includegraphics[width=0.33\textwidth]{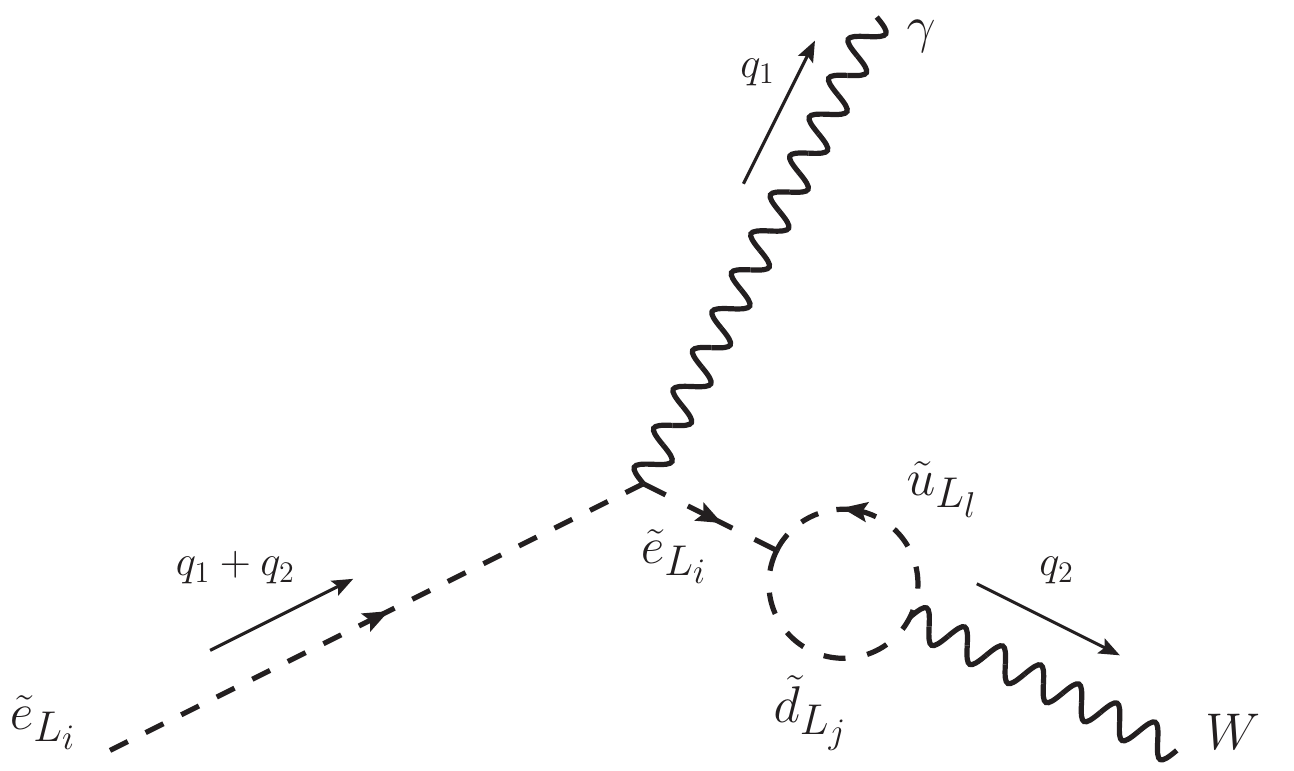}}
\caption{Effective $\tilde{e}_{L} \gamma W$ vertex with squark loops. The corresponding diagrams for slepton loops can be obtained by replacing the up-type squark with a sneutrino and the down-type squark with a charged slepton in the figures. (Note that diagram (a) is absent in that case.)}
\label{fig:1st_loop_sfermion}
\end{figure}
%%%%%%%%%%%%%%%%%%%%%%%%%%%%%%%

In our calculation of the EDM, aside from the differences already discussed in the preceding sections, we find general agreement with Ref.~\cite{Yamanaka:2012ep}. However, while the final expressions are consistent, the treatment of the $W$ boson exchange diagrams involves certain subtleties that require careful attention. To elucidate these aspects, we provide details of our computation for the $W$ boson exchange contributions. Figs.~\ref{fig:1st_loop_fermion} and~\ref{fig:1st_loop_sfermion} illustrate all relevant diagrams contributing to the one-loop effective $\tilde{e}_{L} \gamma W$ vertex with quarks and squarks in the inner loop, including Figs.~\ref{fig:1st_loop_fermion}(e) and~\ref{fig:1st_loop_sfermion}(e) not explicitly reported in Ref.~\cite{Yamanaka:2012ep}. Similar diagrams also arise for lepton and slepton loops, where up-type (s)quarks and down-type (s)quarks in Figs.~\ref{fig:1st_loop_fermion} and~\ref{fig:1st_loop_sfermion} are replaced by (s)neutrinos and charged (s)leptons, respectively. The corresponding expressions for lepton and slepton loops can be obtained by substituting $n_c = 1$, $Q_u = 0$, and $Q_d = -1$, and replacing the internal (s)quark masses with the masses of the relevant (s)leptons appearing in the loop. To facilitate a direct comparison with the calculation procedure in Ref.~\cite{Yamanaka:2012ep}, we adopt the same non-linear $R_{\xi}$ gauge~\cite{Fujikawa:1973qs,Gavela:1981ri,Romao:1986vp}, in which the contribution from Nambu-Goldstone boson ($\phi^-$) exchange, and consequently that of Fig.~\ref{fig:1st_loop_fermion}(d), is absent.

Since Ref.~\cite{Yamanaka:2012ep} provides sufficient derivation details, we present here only the essential steps relevant to our calculation. In the Passarino--Veltman reduction method~\cite{tHooft:1978jhc,Passarino:1978jh,Denner:1991kt}, the one-loop tensor integrals are defined as
{\footnotesize\begin{align}
B_0\left(p_1^2, m_0^2, m_1^2\right) &\equiv \frac{(2 \pi \mu)^\epsilon}{i \pi^2} \int \mathrm{d}^d k \frac{1}{\left[k^2-m_0^2\right]\left[\left(k+p_1\right)^2-m_1^2\right]}~,\\
B^\mu\left(p_1^2, m_0^2, m_1^2\right) &\equiv \frac{(2 \pi \mu)^\epsilon}{i \pi^2} \int \mathrm{d}^d k \frac{k^\mu}{\left[k^2-m_0^2\right]\left[\left(k+p_1\right)^2-m_1^2\right]}~,\\
C_0\left(p_1^2,\left(p_1-p_2\right)^2, p_2^2, m_0^2, m_1^2, m_2^2\right) &\equiv \frac{(2 \pi \mu)^\epsilon}{i \pi^2} \int \mathrm{d}^d k \frac{1}{\left[k^2-m_0^2\right]\left[\left(k+p_1\right)^2-m_1^2\right]\left[\left(k+p_2\right)^2-m_2^2\right]}~,\\
C^\mu\left(p_1^2,\left(p_1-p_2\right)^2, p_2^2, m_0^2, m_1^2, m_2^2\right) &\equiv \frac{(2 \pi \mu)^\epsilon}{i \pi^2} \int \mathrm{d}^d k \frac{k^\mu}{\left[k^2-m_0^2\right]\left[\left(k+p_1\right)^2-m_1^2\right]\left[\left(k+p_2\right)^2-m_2^2\right]}~,\\
C^{\mu \nu}\left(p_1^2,\left(p_1-p_2\right)^2, p_2^2, m_0^2, m_1^2, m_2^2\right) &\equiv \frac{(2 \pi \mu)^\epsilon}{i \pi^2} \int \mathrm{d}^d k \frac{k^\mu k^\nu}{\left[k^2-m_0^2\right]\left[\left(k+p_1\right)^2-m_1^2\right]\left[\left(k+p_2\right)^2-m_2^2\right]}~.
\end{align}}

For the contributions of the diagrams shown in Figs.~\ref{fig:1st_loop_fermion}(a) and~\ref{fig:1st_loop_fermion}(b), aside from certain additional terms that are nonzero at the level of the effective vertex in the first loop but do not contribute to the final EDM, our results are consistent with Ref.~\cite{Yamanaka:2012ep}. Specifically, the effective vertices shown in Figs.~\ref{fig:1st_loop_fermion}(a) and~\ref{fig:1st_loop_fermion}(b) are given by
\begin{align}
    i\mathcal{M}_{\tilde{e} \gamma W}^{(6\mathrm{a})} = -\frac{i e^2 n_c Q_u \lambda^\prime_{inj} V_{lj} V^*_{ln} m_{d_j}}{8 \sqrt{2} \pi^2 s_W} &\big[ ((q_1 \cdot q_2)g^{\mu \nu} - q_2^\mu q_1^\nu) (C_2 + 2C_{12}) \notag \\
    &+ 2 q_2^\mu q_2^\nu C_{22} + i\varepsilon^{\mu \nu \alpha \beta} q_{1 \alpha} q_{2 \beta} C_2 + g^{\mu \nu} B_1 \big]~,
\end{align}
\begin{align}
    i\mathcal{M}_{\tilde{e} \gamma W}^{(6\mathrm{b})} = -\frac{i e^2 n_c Q_d \lambda^\prime_{inj} V_{lj} V^*_{ln} m_{d_j}}{8 \sqrt{2} \pi^2 s_W} &\big[ ((q_1 \cdot q_2)g^{\mu \nu} - q_2^\mu q_1^\nu) (- C_0 - C_2 + 2C_{12}) + 2 q_2^\mu q_2^\nu (C_2 + C_{22}) \notag \\
    & + i\varepsilon^{\mu \nu \alpha \beta} q_{1 \alpha} q_{2 \beta} (C_0 + C_2) + g^{\mu \nu} (B_0 + B_1) \big]~,
\end{align}
where the loop functions are evaluated with the following arguments: ($q_1^2$, $(q_1 + q_2)^2$, $q_2^2$, $m_{u_l}^2$, $m_{u_l}^2$, $m_{d_j}^2$) for the $C$ functions and ($(q_1 + q_2)^2$, $m_{u_l}^2$, $m_{d_j}^2$) for the $B$ functions in $\mathcal{M}_{\tilde{e} \gamma W}^{(6\mathrm{a})}$; and ($q_1^2$, $(q_1 + q_2)^2$, $q_2^2$, $m_{d_j}^2$, $m_{d_j}^2$, $m_{u_l}^2$) and ($(q_1 + q_2)^2$, $m_{d_j}^2$, $m_{u_l}^2$) for the corresponding functions in $\mathcal{M}_{\tilde{e} \gamma W}^{(6\mathrm{b})}$. The momenta $q_1$ and $q_2$ correspond to those labeled in Figs.~\ref{fig:1st_loop_fermion} and \ref{fig:1st_loop_sfermion}, and the meanings of the other variables are the same as in the main text.

In contrast to Ref.~\cite{Yamanaka:2012ep}, our effective vertex corresponding to Fig.~\ref{fig:1st_loop_fermion}(c) exhibits an explicit dependence on the gauge-fixing parameter $\xi$ in the non-linear $R_\xi$ gauge. This is because Fig.~\ref{fig:1st_loop_fermion}(c) involves a $W$-boson propagator, and the $W$ boson connected to the lower fermion line of the Barr--Zee type topology is off shell. As a result, the effective vertex is not gauge-invariant by itself, and the $\xi$-dependence does not vanish at the first-loop level. The amplitude for Fig.~\ref{fig:1st_loop_fermion}(c) reads
\begin{align}
    i\mathcal{M}_{\tilde{e} \gamma W}^{(6\mathrm{c})} = \frac{i e^2 n_c \lambda^\prime_{inj} V_{lj} V^*_{ln} m_{d_j}}{8 \sqrt{2} \pi^2 s_W} \Bigg( \frac{(2q_1 \cdot q_2- (\xi-1) q_2^2) g^{\mu \nu} + (\xi + 1) q_2^\mu q_2^\nu}{(q_1 + q_2)^2 - \xi m_W^2} \Bigg) B_1~,
\end{align}
where the argument of the $B_1$ function in the expression for $\mathcal{M}_{\tilde{e} \gamma W}^{(6\mathrm{c})}$ is ($(q_1 + q_2)^2$, $m_{u_l}^2$, $m_{d_j}^2$).

The divergent $B$-function terms in $\mathcal{M}_{\tilde{e} \gamma W}^{(6\mathrm{a})}$ and $\mathcal{M}_{\tilde{e} \gamma W}^{(6\mathrm{b})}$ cancel out when combined with the contribution from $\mathcal{M}_{\tilde{e} \gamma W}^{(6\mathrm{c})}$, after completing the second-loop integration. Note that in the non-linear $R_\xi$ gauge, the effective vertices $\mathcal{M}_{\tilde{e} \gamma W}^{(6\mathrm{a})}$ and $\mathcal{M}_{\tilde{e} \gamma W}^{(6\mathrm{b})}$ are manifestly $\xi$-independent. In contrast, $\mathcal{M}_{\tilde{e} \gamma W}^{(6\mathrm{c})}$, which involves a $W$-boson propagator, contains explicit $\xi$-dependence. As a result, the cancellation of the divergent terms can only be realized after assembling the full Barr--Zee type diagram with the second-loop integration. Specifically, the final contribution of Fig.~\ref{fig:1st_loop_fermion}(c) to the full Barr--Zee diagram is given by
\begin{align}
    i\mathcal{M}_\mathrm{BZ}^{(6\mathrm{c})} = \frac{e^3 n_c \lambda^\prime_{inj} \lambda^{\prime *}_{ikk} V_{lj} V^*_{ln} m_{d_j}}{8 \pi^2 s_W^2} \int \frac{\mathrm{d}^d q}{(2 \pi)^d} \bar{u}(p^\prime) \gamma^\mu \slashed{q} P_R u(p) \epsilon_\mu^*(q_1) \frac{(q_1 \cdot q) B_1(q^2,m_{u_l}^2,m_{d_j}^2)}{q^2 (q^2 - m_W^2)^2 (q^2 - m^2_{\tilde{e}_{L_i}})}~,
\end{align}
where the spinors $\bar{u}(p^\prime)$ and $u(p)$ refer to the outgoing and incoming external fermions, respectively. Their momenta satisfy $p - p^\prime = q_1 \simeq 0$, reflecting the EDM limit where the external photon is nearly soft.

As expected, the final result from Fig.~\ref{fig:1st_loop_fermion}(c) is free of $\xi$-dependence, and it precisely cancels the divergent $B$-function terms arising from Figs.~\ref{fig:1st_loop_fermion}(a) and \ref{fig:1st_loop_fermion}(b). This cancellation serves as a consistency check of gauge invariance of the two-loop calculation. Note that the expression above corresponds to the Barr--Zee diagram in which the slepton propagator appears on the left-hand side and the $W$-boson propagator is on the right. An analogous cancellation of the divergent terms also occurs in the set of mirror diagrams, where the slepton is located on the right-hand side and the $W$ boson on the left, corresponding to the mirrored versions of Figs.~\ref{fig:1st_loop_fermion}(a), \ref{fig:1st_loop_fermion}(b), and \ref{fig:1st_loop_fermion}(c). The corresponding contribution from the mirror diagram is expressed as
\begin{align}
    i\mathcal{M}_\mathrm{BZ}^{\prime (6\mathrm{c})} = -\frac{e^3 n_c \lambda^{\prime *}_{inj} \lambda^{\prime}_{ikk} V^*_{lj} V_{ln} m_{d_j}}{8 \pi^2 s_W^2} \int \frac{\mathrm{d}^d q}{(2 \pi)^d} \bar{u}(p^\prime) \slashed{q} \gamma^\mu P_L u(p) \epsilon_\mu^*(q_1) \frac{(q_1 \cdot q) B_1(q^2,m_{u_l}^2,m_{d_j}^2)}{q^2 (q^2 - m_W^2)^2 (q^2 - m^2_{\tilde{e}_{L_i}})}~,
\end{align}
where the prime symbol indicates that this amplitude corresponds to the mirrored version, in which the positions of slepton and $W$-boson in the subloops are interchanged.

Since Fig.~\ref{fig:1st_loop_fermion}(d) vanishes in the non-linear $R_\xi$ gauge, and the final contribution from Fig.~\ref{fig:1st_loop_fermion}(e), as will be shown later, is found to vanish as well, the sum of the remaining finite parts of Figs.~\ref{fig:1st_loop_fermion}(a) and~\ref{fig:1st_loop_fermion}(b), together with their mirrored counterparts, collectively yield the expression below after performing some Dirac algebra and applying Gordon-like identities~\cite{Nowakowski:2004cv}:
\begin{align}
i\mathcal{M}_\mathrm{BZ}^{(6\mathrm{a})} + i\mathcal{M}_\mathrm{BZ}^{(6\mathrm{b})} &+ i\mathcal{M}_\mathrm{BZ}^{(6\mathrm{c})} + i\mathcal{M}_\mathrm{BZ}^{\prime (6\mathrm{a})} + i\mathcal{M}_\mathrm{BZ}^{\prime (6\mathrm{b})} + i\mathcal{M}_\mathrm{BZ}^{\prime (6\mathrm{c})} = \notag\\
&-\frac{e \alpha_{\mathrm{em}} n_c \mathrm{Im}(\lambda^\prime_{inj} \lambda^{\prime *}_{ikk} V_{lj} V^*_{ln}) m_{d_j}}{8\pi s_W^2} \bar{u}(p^\prime) \sigma^{\mu \nu} q_{1 \nu} \gamma_5 u(p) \epsilon_\mu^*(q_1) \notag\\
&\times \int \frac{\mathrm{d}^d q}{(2\pi)^d} \int_0^1 \mathrm{d} z \frac{(1-z)Q_u + z Q_d}{\Big(q^2 - \frac{m_{d_j}^2}{1-z} - \frac{m_{u_l}^2}{z}\Big)(q^2 - m_W^2)(q^2 - m^2_{\tilde{e}_{L_i}})}~,
\end{align}
which reproduces the second term in Eq.~\eqref{eq:dk f W}.

Similarly, for the case of squark loops, the contribution from Fig.~\ref{fig:1st_loop_sfermion}(d), which contains explicit $\xi$-dependence at the first loop level, cancels against the divergent $B$-function terms arising from Figs.~\ref{fig:1st_loop_sfermion}(a), \ref{fig:1st_loop_sfermion}(b), and \ref{fig:1st_loop_sfermion}(c) at the level of the full two-loop amplitude. The remaining finite parts of Figs.~\ref{fig:1st_loop_sfermion}(a) and~\ref{fig:1st_loop_sfermion}(b)\footnote{Fig.~\ref{fig:1st_loop_sfermion}(c) does not contribute a finite part beyond the $B$ function terms canceled by Fig.~\ref{fig:1st_loop_sfermion}(d).}, together with their mirrored counterparts, yield the second term in Eq.~\eqref{eq:dk sf W}. Explicitly, the relevant contributions read:
\begin{align}
i\mathcal{M}_{\tilde{e} \gamma W}^{(7\mathrm{a})} = -\frac{i e^2 n_c Q_u \lambda^\prime_{inj} V_{lj} V^*_{ln} m_{d_j}}{8 \sqrt{2} \pi^2 s_W} &\big[ 2((q_1 \cdot q_2)g^{\mu \nu} - q_2^\mu q_1^\nu) C_{12} \notag \\
    &+ q_2^\mu q_2^\nu (C_2 + 2C_{22}) + g^{\mu \nu} (B_0 + B_1) \big]~,
\end{align}
\begin{align}
    i\mathcal{M}_{\tilde{e} \gamma W}^{(7\mathrm{b})} = -\frac{i e^2 n_c Q_d \lambda^\prime_{inj} V_{lj} V^*_{ln} m_{d_j}}{8 \sqrt{2} \pi^2 s_W} &\big[ 2((q_1 \cdot q_2) g^{\mu \nu} - q_2^\mu q_1^\nu) C_{12} \notag \\
    &+ q_2^\mu q_2^\nu (C_2 + 2C_{22}) + g^{\mu \nu} (B_0 + B_1) \big]~,
\end{align}
\begin{align}
    i\mathcal{M}_{\tilde{e} \gamma W}^{(7\mathrm{c})} = \frac{i e^2 n_c (Q_u + Q_d) \lambda^\prime_{inj} V_{lj} V^*_{ln} m_{d_j}}{16 \sqrt{2} \pi^2 s_W} g^{\mu \nu} B_0((q_1+q_2)^2,m_{\tilde{d}_j}^2,m_{\tilde{u}_l}^2)~,
\end{align}
\begin{align}
    i\mathcal{M}_{\tilde{e} \gamma W}^{(7\mathrm{d})} = \frac{i e^2 n_c \lambda^\prime_{inj} V_{lj} V^*_{ln} m_{d_j}}{16 \sqrt{2} \pi^2 s_W} \Bigg( \frac{(2q_1 \cdot q_2- (\xi-1) q_2^2) g^{\mu \nu} + (\xi + 1) q_2^\mu q_2^\nu}{(q_1 + q_2)^2 - \xi m_W^2} \Bigg) (B_0 + 2B_1)~,
\end{align}
\begin{align}
    i\mathcal{M}_\mathrm{BZ}^{(7\mathrm{d})} = \frac{e^3 n_c \lambda^\prime_{inj} \lambda^{\prime *}_{ikk} V_{lj} V^*_{ln} m_{d_j}}{16 \pi^2 s_W^2} \int \frac{\mathrm{d}^d q}{(2 \pi)^d} \bar{u}(p^\prime) \gamma^\mu \slashed{q} P_R u(p) \epsilon_\mu^*(q_1) \frac{(q_1 \cdot q) (B_0 + 2B_1)}{q^2 (q^2 - m_W^2)^2 (q^2 - m^2_{\tilde{e}_{L_i}})}~,
\end{align}
\begin{align}
i\mathcal{M}_\mathrm{BZ}^{(7\mathrm{a})} + i\mathcal{M}_\mathrm{BZ}^{(7\mathrm{b})} &+ i\mathcal{M}_\mathrm{BZ}^{(7\mathrm{c})} + i\mathcal{M}_\mathrm{BZ}^{(7\mathrm{d})} + i\mathcal{M}_\mathrm{BZ}^{\prime (7\mathrm{a})} + i\mathcal{M}_\mathrm{BZ}^{\prime (7\mathrm{b})} + i\mathcal{M}_\mathrm{BZ}^{\prime (7\mathrm{c})} + i\mathcal{M}_\mathrm{BZ}^{\prime (7\mathrm{d})} = \notag\\
&\frac{e \alpha_{\mathrm{em}} n_c \mathrm{Im}(\lambda^\prime_{inj} \lambda^{\prime *}_{ikk} V_{lj} V^*_{ln}) m_{d_j}}{8\pi s_W^2} \bar{u}(p^\prime) \sigma^{\mu \nu} q_{1 \nu} \gamma_5 u(p) \epsilon_\mu^*(q_1) \notag\\
&\times \int \frac{\mathrm{d}^d q}{(2\pi)^d} \int_0^1 \mathrm{d} z \frac{(1-z)Q_u + z Q_d}{\Big(q^2 - \frac{m_{\tilde{d}_j}^2}{1-z} - \frac{m_{\tilde{u}_l}^2}{z}\Big)(q^2 - m_W^2)(q^2 - m^2_{\tilde{e}_{L_i}})}~,
\end{align}
where the arguments of the loop functions are specified as follows: for $\mathcal{M}_{\tilde{e} \gamma W}^{(7\mathrm{a})}$, the $C$ functions depend on ($q_1^2$, $(q_1 + q_2)^2$, $q_2^2$, $m_{\tilde{u}_l}^2$, $m_{\tilde{u}_l}^2$, $m_{\tilde{d}_j}^2$), and the $B$ functions on ($(q_1 + q_2)^2$, $m_{\tilde{u}_l}^2$, $m_{\tilde{d}_j}^2$); for $\mathcal{M}_{\tilde{e} \gamma W}^{(7\mathrm{b})}$, the $C$ functions depend on ($q_1^2$, $(q_1 + q_2)^2$, $q_2^2$, $m_{\tilde{d}_j}^2$, $m_{\tilde{d}_j}^2$, $m_{\tilde{u}_l}^2$), and the $B$ functions on ($(q_1 + q_2)^2$, $m_{\tilde{d}_j}^2$, $m_{\tilde{u}_l}^2$); for $\mathcal{M}_{\tilde{e} \gamma W}^{(7\mathrm{c})}$, the $B_0$ function depends on ($(q_1 + q_2)^2$, $m_{\tilde{d}_j}^2$, $m_{\tilde{u}_l}^2$), and the $B_1$ function on ($(q_1 + q_2)^2$, $m_{\tilde{u}_l}^2$, $m_{\tilde{d}_j}^2$); and for $\mathcal{M}_\mathrm{BZ}^{(7\mathrm{d})}$, the $B_0$ function depends on ($q^2$, $m_{\tilde{d}_j}^2$, $m_{\tilde{u}_l}^2$), and the $B_1$ function on ($q^2$, $m_{\tilde{u}_l}^2$, $m_{\tilde{d}_j}^2$).

Finally, as mentioned earlier, it turns out that the contributions from Figs.~\ref{fig:1st_loop_fermion}(e) and~\ref{fig:1st_loop_sfermion}(e) vanish, despite the one-loop effective $\tilde{e}_{L} \gamma W$ vertex being non-zero. Specifically, the effective vertex corresponding to Fig.~\ref{fig:1st_loop_fermion}(e) takes the form:
\begin{align}
    i\mathcal{M}_{\tilde{e} \gamma W}^{(6\mathrm{e})} = -\frac{i e^2 n_c \lambda^\prime_{inj} V_{lj} V^*_{ln} m_{d_j}}{4 \sqrt{2} \pi^2 s_W} \frac{q_2^{\mu} q_2^{\nu}}{q_2^2 - m^2_{\tilde{e}_{L_i}}} B_1(q_2^2,m_{u_l}^2,m_{d_j}^2)~.
\end{align}
The contribution of this term to the second-loop Barr--Zee type diagram is given by
\begin{align}
    i\mathcal{M}_\mathrm{BZ}^{(6\mathrm{e})} = \frac{e^3 n_c \lambda^\prime_{inj} \lambda^{\prime *}_{ikk} V_{lj} V^*_{ln} m_{d_j}}{4 \pi^2 s_W^2} \bar{u}(p^\prime) P_R u(p) \epsilon_\mu^*(q_1) \int \frac{\mathrm{d}^d q}{(2 \pi)^d} \frac{(q_1 \cdot q) q^\mu B_1(q^2,m_{u_l}^2,m_{d_j}^2)}{(q^2 / \xi - m_W^2) (q^2 - m^2_{\tilde{e}_{L_i}})^3}~.
\end{align}
This contribution vanishes upon loop integration as a result of the transversality condition $\epsilon_{\mu} q_1^\mu = 0$.

As for the effective vertex associated with Fig.~\ref{fig:1st_loop_sfermion}(e):
\begin{align}
    i\mathcal{M}_{\tilde{e} \gamma W}^{(7\mathrm{e})} = \frac{i e^2 n_c \lambda^\prime_{inj} V_{lj} V^*_{ln} m_{d_j}}{8 \sqrt{2} \pi^2 s_W} \frac{q_2^{\mu} q_2^{\nu}}{q_2^2 - m^2_{\tilde{e}_{L_i}}} \big[B_0(q_2^2,m_{\tilde{d}_j}^2,m_{\tilde{u}_l}^2) + 2 B_1(q_2^2,m_{\tilde{u}_l}^2,m_{\tilde{d}_j}^2)\big]~.
\end{align}
Accordingly, the corresponding contribution to the second-loop Barr--Zee diagram is:
\begin{align}
    i\mathcal{M}_\mathrm{BZ}^{(7\mathrm{e})} &= -\frac{e^3 n_c \lambda^\prime_{inj} \lambda^{\prime *}_{ikk} V_{lj} V^*_{ln} m_{d_j}}{8 \pi^2 s_W^2} \bar{u}(p^\prime) P_R u(p) \epsilon_\mu^*(q_1) \notag\\
    &\times \int \frac{\mathrm{d}^d q}{(2 \pi)^d} \frac{(q_1 \cdot q) q^\mu \big[B_0(q^2,m_{\tilde{d}_j}^2,m_{\tilde{u}_l}^2) + 2 B_1(q^2,m_{\tilde{u}_l}^2,m_{\tilde{d}_j}^2)\big]}{(q^2 / \xi - m_W^2) (q^2 - m^2_{\tilde{e}_{L_i}})^3}~.
\end{align}
This term similarly vanishes by the same reasoning as for $\mathcal{M}_\mathrm{BZ}^{(6\mathrm{e})}$. The same conclusion applies to $\mathcal{M}_\mathrm{BZ}^{\prime (6\mathrm{e})}$ and $\mathcal{M}_\mathrm{BZ}^{\prime (7\mathrm{e})}$, whose contributions also vanish in the complete two-loop amplitude.

\end{appendix}

%%%%%%%%%%%%%%%%%%%%%%%%%%%%%%%%%%%
\bibliographystyle{JHEP}
\bibliography{EDMs}

\end{document}